\documentclass[useAMS,usenatbib]{mn2e}
\usepackage{graphicx}
\usepackage{times}
\usepackage{wrapfig}
\makeatletter

\newcommand{\Rmnum}[1]{\expandafter\@slowromancap\romannumeral #1@}
\makeatother
\title[GAMA spectroscopic analysis]{Galaxy And Mass Assembly (GAMA): Spectroscopic analysis}

\author[A. M. Hopkins et al.]
{A. M. Hopkins$^{1}$\thanks{E-mail:ahopkins@aao.gov.au},
S. P. Driver$^{2,3}$, S. Brough$^{1}$, M. S. Owers$^{1}$, A. E. Bauer$^{1}$,
\newauthor
 M. L. P. Gunawardhana$^{1,4}$, M. E. Cluver$^{1}$, M. Colless$^{1}$, C. Foster$^{5}$, M. A. Lara-L{\'o}pez$^{1}$,
\newauthor
I. Roseboom$^{6}$, R. Sharp$^{7}$, O. Steele$^{8}$, D. Thomas$^{8}$,
I. K. Baldry$^{9}$, M. J. I. Brown$^{10}$, J. Liske$^{11}$,
\newauthor
P. Norberg$^{12}$, A. S. G. Robotham$^{2,3}$, S. Bamford$^{13}$,
J. Bland-Hawthorn$^{4}$,  M. J. Drinkwater$^{14}$,
\newauthor
J. Loveday$^{15}$, M. Meyer$^{2}$, J. A. Peacock$^{6}$, R. Tuffs$^{16}$,
N. Agius$^{17}$, M. Alpaslan$^{2,3}$, E. Andrae$^{16}$,
\newauthor
E. Cameron$^{16}$, S. Cole$^{12}$, J. H. Y. Ching$^{4}$,
L. Christodoulou$^{15}$, C. Conselice$^{13}$, S. Croom$^{4}$,
\newauthor
N. J. G. Cross$^{6}$, R. De Propris$^{18}$, J. Delhaize$^{2}$, L. Dunne$^{19}$, 
S. Eales$^{20}$, S. Ellis$^{1}$, C. S. Frenk$^{12}$,
\newauthor
A. Graham$^{21}$, M. W. Grootes$^{16}$, B. H{\"a}u{\ss}ler$^{13}$,
C. Heymans$^{6}$, D. Hill$^{3}$, B. Hoyle$^{22}$, M. Hudson$^{23}$,
\newauthor
 M. Jarvis$^{24,25}$, J. Johansson$^{26}$, D. H. Jones$^{10}$, E. van Kampen$^{11}$, L. Kelvin$^{2,3}$,
 K. Kuijken$^{27}$,
\newauthor
{\'A}. L{\'o}pez-S{\'a}nchez$^{1,28}$, S. Maddox$^{19}$,
B. Madore$^{29}$, C. Maraston$^{8}$, T. McNaught-Roberts$^{12}$,
\newauthor
R. C. Nichol$^{8}$, S. Oliver$^{15}$,
H. Parkinson$^{6}$, S. Penny$^{10}$, S. Phillipps$^{30}$, K. A. Pimbblet,$^{10}$, 
\newauthor
T. Ponman$^{31}$, C. C. Popescu$^{17}$, M. Prescott$^{25}$, R. Proctor$^{32}$,
E. M. Sadler$^{4}$,
A. E. Sansom$^{17}$,
\newauthor
M. Seibert$^{29}$, L. Staveley-Smith$^{2}$, W. Sutherland$^{33}$, E. Taylor$^{4}$, 
L. Van Waerbeke$^{34}$,
\newauthor
J. A. V{\'a}zquez-Mata$^{15}$, S. Warren$^{35}$,
D. B. Wijesinghe$^{4}$, V. Wild$^{3}$, S. Wilkins$^{24}$ \\ 
$^{1}$Australian Astronomical Observatory, PO Box 915, North Ryde, NSW 1670, Australia\\
$^{2}$International Centre for Radio Astronomy Research (ICRAR), 
University of Western Australia, Crawley, WA 6009, Australia \\
$^{3}$School of Physics \& Astronomy, University of St Andrews, North Haugh, St Andrews, KY16 9SS, UK\\
$^{4}$Sydney Institute for Astronomy, School of Physics, University of Sydney, NSW 2006, Australia\\
$^{5}$European Southern Observatory, Alonso de Cordova 3107, Vitacura, Santiago, Chile\\
$^{6}$Institute for Astronomy, University of Edinburgh, Royal Observatory, Blackford Hill, Edinburgh, EH9 3HJ, UK\\
$^{7}$Research School of Astronomy \& Astrophysics, Australian National University, Cotter Road
Weston Creek, ACT 2611, Australia\\
$^{8}$Institute of Cosmology and Gravitation (ICG), University of Portsmouth, Dennis Sciama Building, Burnaby Road, Portsmouth PO1 3FX, UK\\
$^{9}$Astrophysics Research Institute, Liverpool John Moores University, Twelve Quays House, Egerton Wharf, Birkenhead, CH41 1LD, UK\\
$^{10}$School of Physics, Monash University, Clayton, Victoria 3800, Australia\\
$^{11}$European Southern Observatory, Karl-Schwarzschild-Str.~2, 85748, Garching, Germany\\
$^{12}$Institute for Computational Cosmology, Department of Physics, Durham University, South Road, Durham, DH1 3LE, UK\\
$^{13}$School of Physics \& Astronomy, University of Nottingham, University Park, Nottingham NG7 2RD, UK\\
$^{14}$School of Physics, University of Queensland, Brisbane QLD 4072, Australia\\
$^{15}$Astronomy Centre, University of Sussex, Falmer, Brighton BN1 9QH, UK\\
$^{16}$Max Planck Institute for Nuclear Physics (MPIK), Saupfercheckweg 1, 69117, Heidelberg, Germany\\
$^{17}$Jeremiah Horrocks Institute, University of Central Lancashire, Preston PR1 2HE, UK\\
$^{18}$Cerro Tololo Inter-American Observatory, La Serena, Chile\\
$^{19}$Department of Physics and Astronomy, University of Canterbury, Private Bag 4800, Christchurch, NZ\\
$^{20}$School of Physics and Astronomy, Cardiff University, The Parade, Cardiff CF24 3AA, UK\\
$^{21}$Centre for Astrophysics \& Supercomputing, Swinburne University, Hawthorn, VIC 3122, Australia\\
$^{22}$Institut de Ci{\`e}ncies del Cosmos, Facultat de F{\'i}sica, Mart{\'i} i Franqu{\`e}s 1, E-08028 Barcelona, Spain\\
$^{23}$Department of Physics \& Astronomy, University of Waterloo, Waterloo, ON N2L 1Z5, Canada\\
$^{24}$University of Oxford, Department of Physics, Denys Wilkinson Building, Keble Road, Oxford, OX1 3RH, UK\\ 
$^{25}$Physics Department, University of the Western Cape, Private Bag X17, Bellville 7535, South Africa\\
$^{26}$Max-Planck Institut fuer Astrophysik, Karl-Schwarzschild-Str.\ 1, 85741 Garching, Germany\\
$^{27}$Leiden Observatory, Leiden University, PO Box 9513, 2300RA Leiden, The Netherlands\\
$^{28}$Department of Physics and Astronomy, Macquarie University, NSW 2109, Australia\\
$^{29}$Carnegie Institution for Science, 813, Santa Barbara Street, Pasadena, California, 91101, USA\\
$^{30}$Department of Physics, University of Bristol, Tyndal l Avenue, Bristol BS8 1TL, UK \\
$^{31}$School of Physics and Astronomy, University of Birmingham, Edgbaston, Birmingham, B15 2TT, UK\\
$^{32}$Universidade de S{\~a}o Paulo, IAG, Rua do Mato 1226, S{\~a}o Paulo 05508-900, Brazil\\
$^{33}$Astronomy Unit, Queen Mary University London, Mile End Rd, London E1 4NS, UK\\
$^{34}$University of British Columbia, Vancouver, BC V6T 2C2, Canada\\
$^{35}$Astrophysics Group, Imperial College London, Blackett Laboratory, Prince Consort Road, London SW7 2AZ, UK\\
}
\begin{document}

\date{Accepted 2013 January 4}

\pagerange{\pageref{firstpage}--\pageref{lastpage}} \pubyear{2013}

\maketitle

\label{firstpage}

\clearpage
\begin{abstract}
The Galaxy And Mass Assembly (GAMA) survey is a multiwavelength photometric and spectroscopic survey,
using the AAOmega spectrograph on the Anglo-Australian Telescope to obtain spectra for up to
$\sim\,300\,000$ galaxies over 280 square degrees, to a limiting magnitude of $r_{\rm pet}<19.8$\,mag.
The target galaxies are distributed over $0<z \la 0.5$ with a median redshift of $z\approx 0.2$, although the redshift
distribution includes a small number of systems, primarily quasars, at higher redshifts, up to and beyond $z=1$.
The redshift accuracy ranges from $\sigma_v\approx 50\,$km\,s$^{-1}$ to $\sigma_v\approx 100\,$km\,s$^{-1}$
depending on the signal-to-noise of the spectrum.
Here we describe the GAMA spectroscopic reduction and analysis pipeline. We present the steps involved in
taking the raw two-dimensional spectroscopic images through to flux-calibrated one-dimensional spectra.
The resulting GAMA spectra cover an observed wavelength range of $3750 \la \lambda \la 8850\,$\AA\ at
a resolution of $R\approx1300$. The final flux calibration is typically accurate
to $10-20\%$, although the reliability is worse at the extreme wavelength ends, and poorer in the blue
than the red.
We present details of the measurement of emission and absorption features in the GAMA spectra.
These measurements are characterised
through a variety of quality control analyses detailing the robustness and reliability of the
measurements. We illustrate the quality of the measurements with a brief exploration of
elementary emission line properties of the galaxies in the GAMA sample. We demonstrate
the luminosity dependence of the Balmer decrement, consistent with previously published
results, and explore further how Balmer decrement varies with galaxy mass and redshift. We
also investigate the mass and redshift dependencies of the [NII]/H$\alpha$ vs [OIII]/H$\beta$
spectral diagnostic diagram, commonly used to discriminate between star forming and
nuclear activity in galaxies.
\end{abstract}

\maketitle

\section{Introduction}
Galaxy surveys that have moderate resolution ($R\sim 1000-2000$) optical spectroscopic
measurements, such as the Two-degree Field Galaxy Redshift Survey \citep[2dF;][]{Col:01} and the
Sloan Digital Sky Survey \citep[SDSS;][]{Yor:00} are among the most productive resources
for understanding galaxy formation
and evolution. The astrophysical information encoded in the rest-frame optical region of the
spectrum is among the most well-understood and well-calibrated aspect of galaxy evolution
studies, and provides a wealth of detail regarding the physical processes occurring within galaxies.
This ranges from quantitative measurements of star formation rate (SFR), metallicity, velocity
dispersion, obscuration, and more, through to  diagnostics distinguishing between the presence
of star formation or an accreting central supermassive black hole (an active galactic nucleus, AGN).
In combination with broadband photometric measurements
spanning ultraviolet through to radio wavelengths, the redshift and other physical
information from galaxy spectra provides a powerful tool for exploring the details of
galaxy evolution.

The Galaxy And Mass Assembly (GAMA)\footnote{\tt http://www.gama-survey.org/}
survey is a large multiwavelength photometric
and spectroscopic galaxy survey \citep{Dri:09,Dri:11} that provides exactly this comprehensive
selection of photometric and spectroscopic data. The key scientific goals are to use the galaxy
distribution to conduct a series of tests of the cold dark matter (CDM) paradigm, in addition to
carrying out detailed studies of the internal structure and evolution of the galaxies themselves.
The scientific motivation for GAMA, the survey footprint, data processing, catalogue construction
and quality control are described by \citet{Dri:11}. The target selection,
including survey masks, star-galaxy separation, and target prioritisation is
presented by \citet{Bal:10}, with the tiling algorithm described by \citet{Rob:10} and
the photometric analysis by \citet{Hill:11}. Stellar masses for the GAMA galaxies have been
quantified by \citet{Tay:11}, and the low redshift stellar mass function is detailed in \citet{Bal:12}.
The broadband luminosity functions are derived by \citet{Lov:12}, and the H$\alpha$ luminosity
functions and evolution presented in \citet{Gun:13}. Galaxy nebular metallicity measurements are
detailed in \citet{Fos:12} and \citet{LL:13}. Galaxy groups in GAMA have been quantified by
\citet{Rob:11}, and galaxy structural parameters measured by \citet{Kel:12}.

The GAMA survey used 68 nights of observing time on the Anglo-Australian
Telescope (AAT) over 2008--2010. This time was used to conduct a highly complete survey
in three Equatorial fields to $r_{\rm pet}<19.4$\,mag over a total of 144\,deg$^2$,
48\,deg$^2$ of which was observed to the deeper limit of $r_{\rm pet}<19.8$\,mag.
This initial phase of the survey, usually referred to as ``GAMA I," allowed the acquisition of
over 112\,000 new galaxy spectra
and redshifts, for a total of over 130\,000 redshifts in the original GAMA survey area.
Subsequently the survey has been extended, with the award of 110 nights of AAT time over
2010--2012, to expand the survey by including two Southern fields and broadening the
three Equatorial fields (referred to as ``GAMA II"). This expands the total survey area to 280\,deg$^2$,
while achieving a uniform depth of $r_{\rm pet}<19.8$\,mag over the full survey region. The goal is to
compile $\sim 300\,000$ galaxy spectra over this area. At the time of writing, we
have already obtained over 220\,000 spectra. In detail, to date GAMA has observed
224\,465 spectra of which 222\,294 are galaxy targets. These numbers include repeat observations,
and not all are main survey targets, as they include ``filler" targets that take advantage of
fibres unable to be allocated to main survey targets on any given observation plate \cite[see][]{Bal:10}.
Including spectra from other surveys within the GAMA regions,
\citep[SDSS, 2dFGRS and others, see][for details]{Bal:10}, we have 299\,980 spectra, of
which 297\,067 are of galaxy targets (not all are main survey targets). We have 233\,777
unique galaxy targets, of which 215\,458 (92.2\%) have redshift quality $nQ\ge3$
\citep[see][for definition of $nQ$, but in brief $nQ=3$ or $nQ=4$ correspond to reliable redshifts]{Dri:11}.

The survey has already led to a number of published results making use of the
detailed emission and absorption line measurements from the GAMA spectroscopic
data. These include identification of
the lowest-mass star forming galaxy population \citep{Bro:11}, a self-consistent approach
to galaxy star formation rate estimates and the role of obscuration \citep{Wij:11a,Wij:11b},
and evidence for a star formation rate dependence in the high-mass slope of the
stellar initial mass function \citep{Gun:11}, among many other GAMA team publications
(for a full and current list see the team web page),
along with additional work led by collaborating surveys such as
{\em Herschel}-ATLAS\footnote{\tt http://www.h-atlas.org/}.

In order to best facilitate subsequent scientific analyses of public survey data, it is crucial
to provide full details of the observations, processing and data product derivations \citep[e.g.,][]{Bol:12}.
Here we describe the spectroscopic pipeline processing for the GAMA survey. This
encompasses an overview of the observations (\S\,\ref{obs}), the steps involved in processing the
raw two-dimensional spectroscopic images through to extracted one-dimensional spectra and
the initial redshift measurement process (\S\,\ref{2dfdr}), and flux calibration
of the one-dimensional spectra (\S\,\ref{corrections}). We also present the processes
used in measuring the emission and absorption features of the spectra (\S\,\ref{measure})
that are recorded in the GAMA database and made publicly available through the
staged data releases. Note that the GAMA spectroscopic reduction and analysis pipeline
is still evolving as we continue to improve some aspects. Here we describe the pipeline
that was used to construct the final GAMA I dataset. It was this dataset that has been used
in the in the various investigations cited above. The spectra and associated measurements
that will be available in GAMA DR2, the public data release due in January 2013, also
rely on the pipeline described here. For the purposes of this paper we use the data
associated with GAMA SpecCat v08.

Throughout, all magnitudes are given in the AB system, and we assume a cosmology with
$H_{0}=$70\,km\,s$^{-1}$\,Mpc$^{-1}$, $\Omega_{M}=0.3$ and $\Omega_{\Lambda}=0.7$.

\begin{figure*}
\centerline{\includegraphics[width=0.7\textwidth]{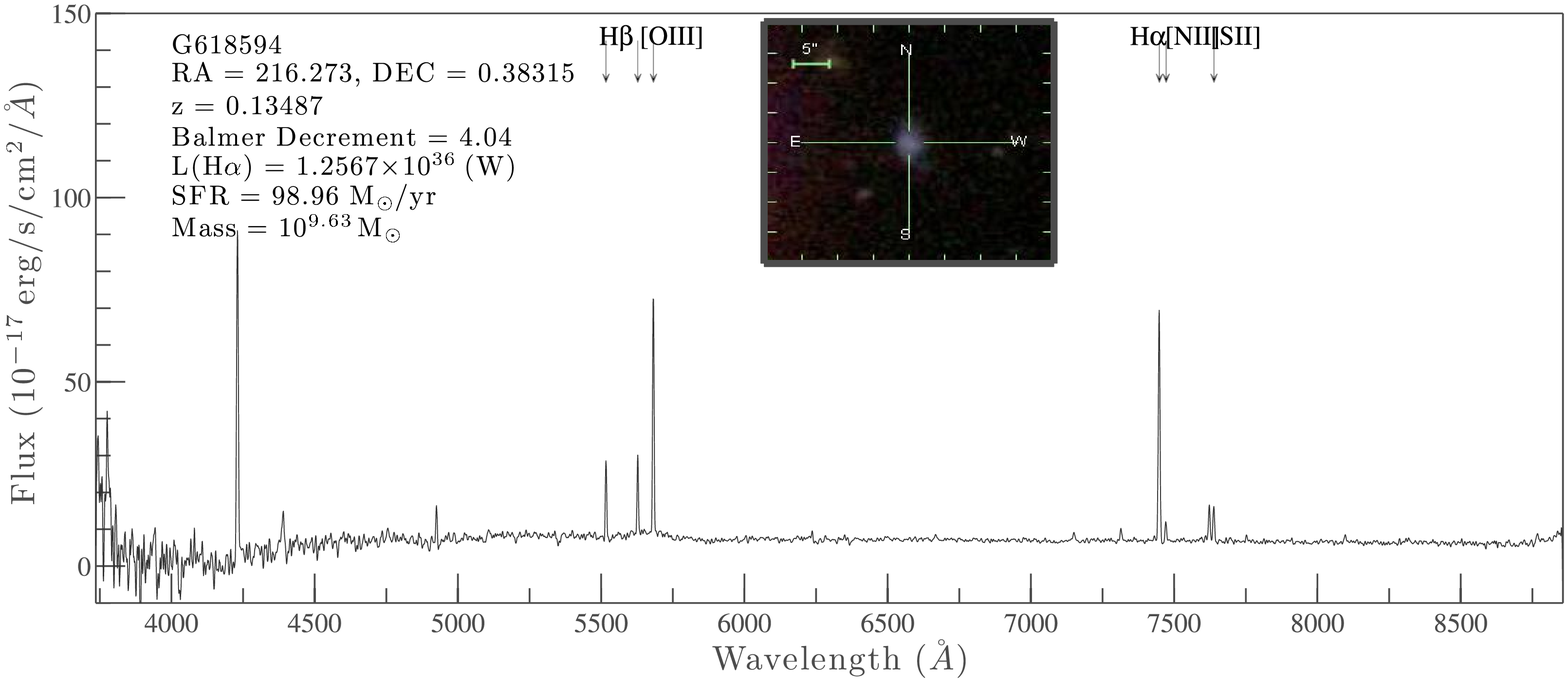}}
\centerline{\includegraphics[width=0.7\textwidth]{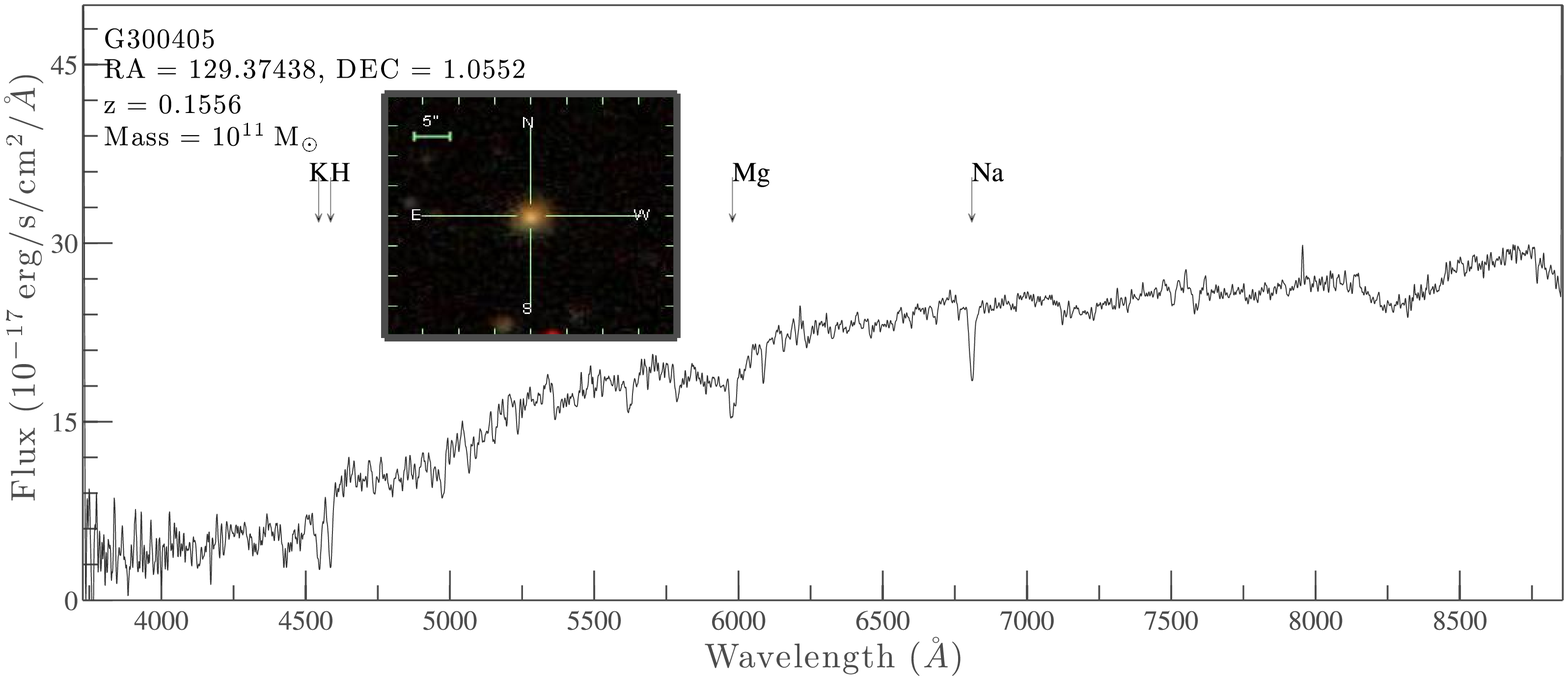}}
\centerline{\includegraphics[width=0.7\textwidth]{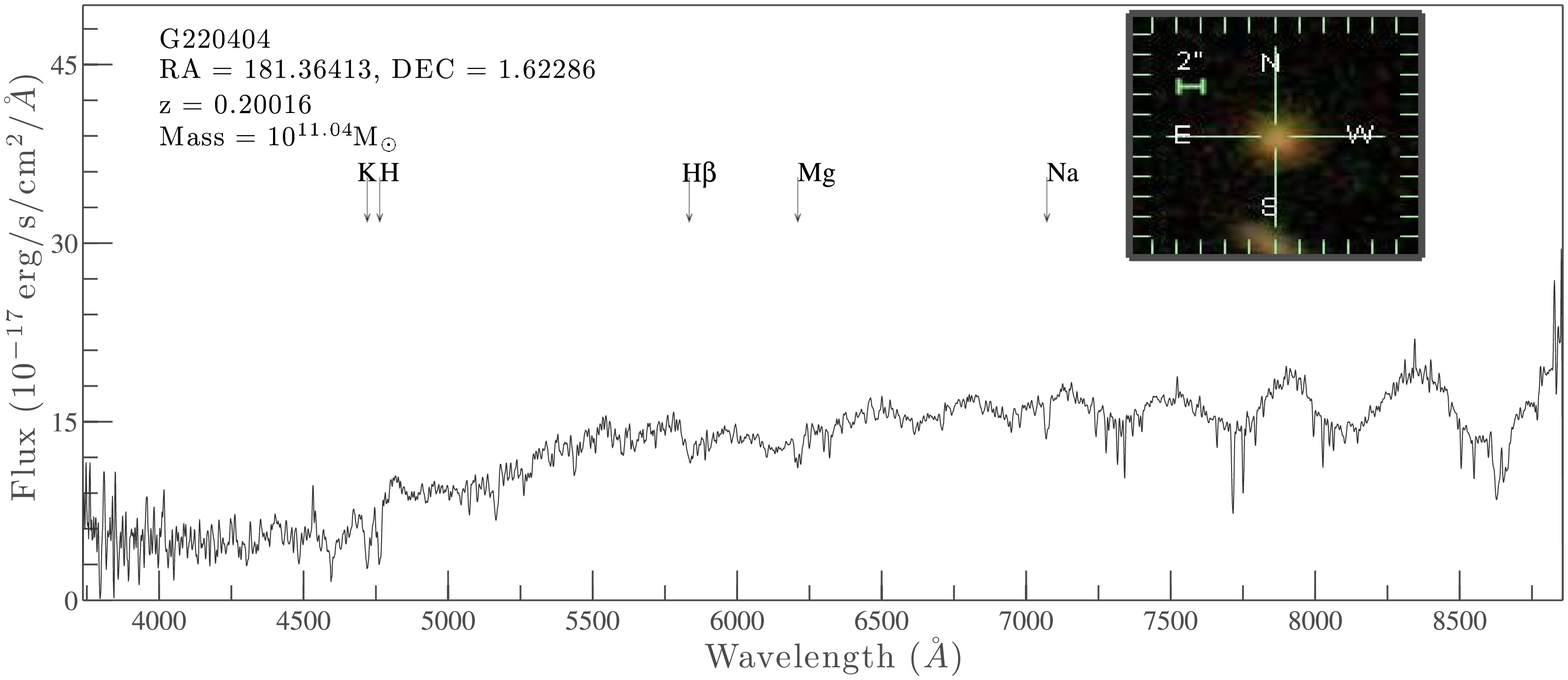}}
\centerline{\includegraphics[width=0.7\textwidth]{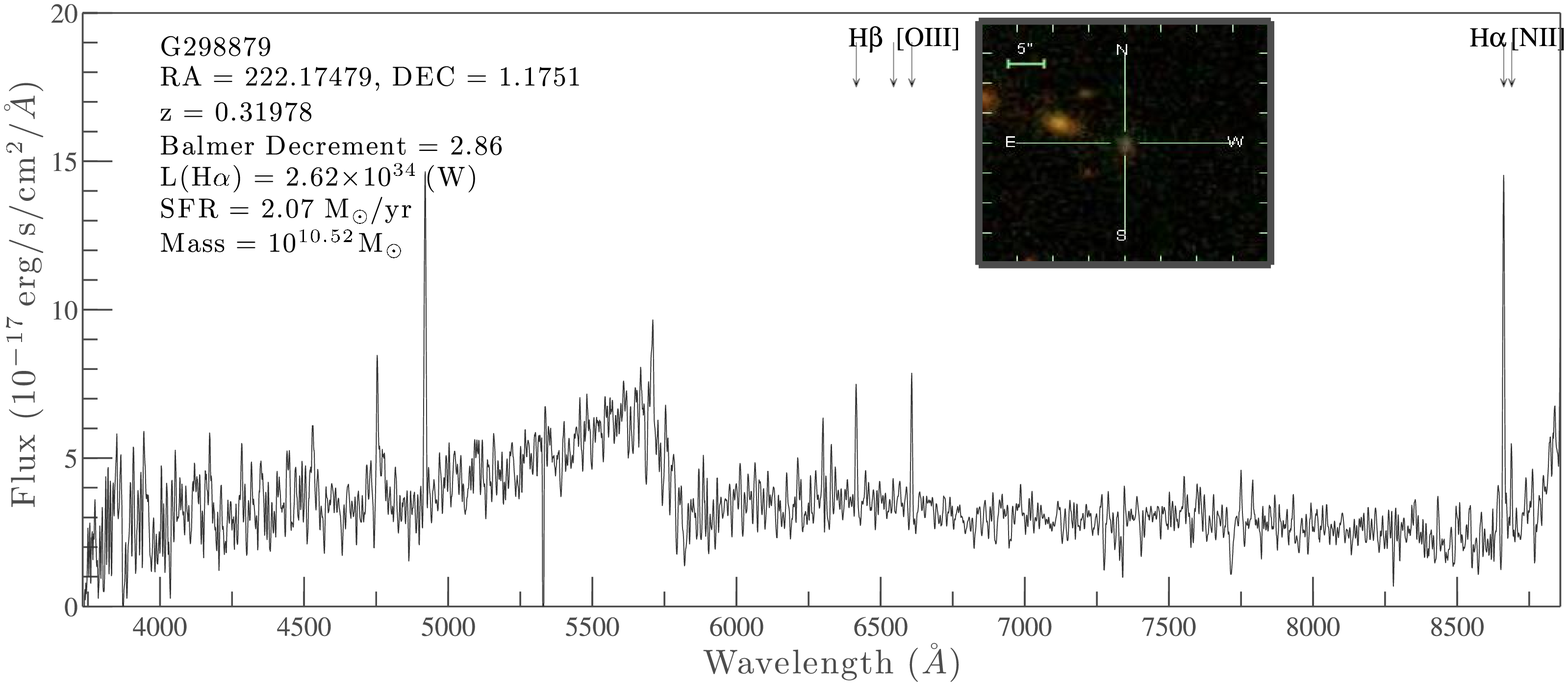}}
\caption{Example GAMA spectra for a selection of galaxies, illustrating high
quality spectra and spectra affected by instrumental and processing artifacts. The spectra
have been smoothed with a five-pixel running boxcar average to aid in clarity of display. From
top to bottom: Star forming galaxy spectrum; Absorption line galaxy spectrum;
Spectrum affected by fringing; Spectrum affected by bad splicing. Each spectrum includes
an inset showing the SDSS colour galaxy image, as well as identifying common emission
or absorption features. Each galaxy has its GAMA ID, redshift and stellar mass \citep[from][]{Tay:11}
listed, along with Balmer decrement, H$\alpha$ luminosity and star formation rate for
star forming galaxy spectra. Only about 3\% of GAMA spectra are affected by fringing or bad splicing.
\label{fig:spectrum}}
\end{figure*}

\section{Observations at the AAT}
\label{obs}

The GAMA spectroscopic observations use the AAOmega spectrograph \citep{Sau:04,Smi:04,Sha:06}
on the 3.9\,m Anglo-Australian Telescope (Siding Spring Observatory, NSW, Australia) for measuring the spectra
of the target galaxies. This spectrograph is stationed in the thermally stable environment
of one of the telescope Coud{\'e} rooms. AAOmega possesses a dual beam system
which allows coverage of the wavelength range from 3750\,\AA\ to 8850\,\AA\
with the 5700\,\AA\ dichroic used by GAMA, in a single observation. Each arm of the
AAOmega system is equipped with a 2k$\times$4k E2V CCD detector and an AAO2 CCD controller.
The blue arm CCD is thinned for improved blue response. The red arm CCD
is a low fringing type. The grating used in the blue arm is the 580V,
centred at 4800\,\AA, which has a dispersion of 1\,\AA/pixel and gives a coverage
of 2100\,\AA. The 385R grating is used in the red arm, centred at 7250\,\AA.
This grating has a dispersion of 1.6\,\AA/pixel and gives a coverage of
3200\,\AA. This leads to spectra with a resolution that varies as
a function of wavelength, from $R\approx 1000$ at the blue end up to $R\approx 1600$ at
the red end.

The ``Two-degree Field" (2dF) instrument \citep{Lew:02} consists of a
wide field corrector, an atmospheric dispersion compensator (ADC), and a robot
gantry which positions optical fibres to an accuracy of $0\farcs3$ on the sky.  The fibres
have a $2''$ diameter projected on the sky \citep{Lew:02}. A tumbling
mechanism with two field plates allows the next field to be configured
while the current field is being observed. The 392 target fibres from 2dF are
fed to the AAOmega spectrograph, and eight guide fibre-bundles
are used to ensure accurate telescope positioning over the course of each exposure.
For GAMA observations, the 2dF robot fibre positioner is
used to configure typically 345 fibres to observe galaxies within a two
degree field on the sky. Due to a varying number of damaged or unusable fibres (typically around 20),
the actual number of fibres able to be used for galaxy targets is not constant. To quantify this,
the first-quartile/median/third-quartile number of fibres on galaxy targets for GAMA I was
332/345/348. For the full survey to date, these numbers are 324/341/348.
This distribution has remained fairly steady over the duration of the survey so far.
Around 25 additional fibres are used to measure the sky spectrum in each field. Sky positions
were identified using a sky mask, detailed in \citet{Bal:10}. Three fibres are allocated
to spectroscopic standard stars. 

The integration time for each GAMA field is typically 60\,min, split into
three 1200\,s exposures. Accounting for the read-out time of the CCDs
(2\,min) and the acquisition of the calibration frames comprising flat-fields and
arc-line exposures for wavelength calibration, the time spent on each field
is well-matched to the time required for the 2dF positioner to configure the
following observing plate, ensuring an efficient overall survey strategy.
Between 20 and 30 bias frames are taken during each observing session.
Starting in 2011, we also began to use dark frames to refine the calibration,
with from 10 to 30 dark frames, each of 1200\,s exposure, being acquired each
observing session.

\section{Data processing and redshift measurements}
\label{2dfdr}

\subsection{Obtaining 1D spectra}
\label{1Dspec}

The raw data are processed using software developed at the AAO
called 2{\sc dfdr}, \citep{CSH:04,SB:10}.
The 2{\sc dfdr} processing applies the standard sequence of tasks for 1D spectral
extraction from 2D images. This includes bias subtraction, flat-fielding, fibre trace (or ``tramline") fitting,
and wavelength calibration. First, a master bias (and for GAMA II data a master dark)
are created using the available bias (and dark) frames. For each new plate
configuration observed, the raw AAOmega frames
are run through 2{\sc dfdr} at the telescope to provide the processed spectra.

The standard parameters are used in running 2{\sc dfdr}, with the following
modifications. We consider not only the master bias (and master dark), but also an overscan
correction using a ninth-order polynomial fit for the blue spectra, and second-order for the red spectra.
The high order is necessary in the blue due to a very strong gradient in the first $\sim 100$ pixels
that is not well-modelled by lower order fits. The wavelength solution is determined by a third-order
polynomial fit to the arc-lines, with the solution tweaked using a first-order (blue) or third-order (red)
polynomial to the sky lines. The higher order is required for the red CCD as there are more sky lines
and the sky is brighter at the red end of the spectrum. The wavelength calibration is referenced to arc
line wavelengths in air rather than vacuum, contrary to the convention adopted by SDSS, and is accurate to
better than $0.1\,$\AA, as measured from key strong sky line features.
Cosmic rays are cleaned in each object frame using
an implementation of Laplacian cosmic ray identification \citep{vD:01}, and applying clipping thresholds of
$10\,\sigma$ in the blue and $5\,\sigma$ in the red. The scattered light is subtracted assuming a
first-order polynomial fit. The throughput calibration method considers a flux weighted value of the night sky
emission lines to normalise the fibre throughputs.

Extraction of spectra is performed by first identifying fibres in a flat field frame and
then fitting their locations as a function of CCD position using a model of the spectrograph
optical distortion. Once the tramlines are located, a minimum variance
Gaussian weighted extraction \citep{SB:10,Hor:86} is used to
obtain the flux in each fibre per spectral pixel. While optimally weighted, this does not take
into account cross-talk between fibres, but given the restricted dynamic range of the GAMA targets
(a range of less than three magnitudes) the level of cross-talk has been shown to be negligible
\citep[their Figure~2]{SB:10}. For each 1D spectrum there exists a variance array determined
from the Poisson noise in the bias corrected 2D frame, the read noise and the gain, and propagated
through the reduction pipeline. Examination of repeat spectra shows that the uncertainties per pixel
are well characterised by the measurements in the variance array, except around the strong $5577\,$\AA\
sky line, and at the extreme ends of the wavelength range. Where there are differences
between the repeat measurements and the variance arrays due to such systematics,
the differences are always less than a factor of $1.4$.

The initial sky subtraction is performed using the $\sim$25 fibres allocated in
each plate to sky positions. A combined sky spectrum is made by taking the
median of the corresponding pixels in each of the normalised sky fibres, discarding the two
brightest sky fibres to avoid potential problems in the event of inadvertent non-sky flux (due
to an asteroid, passing satellite, or other moving object perhaps), at the sky fibre location.
The continuum sky subtraction accuracy is typically 2-3\%\ of the sky level, although for
especially strong sky lines such as that at $5577\,$\AA, the residuals can be worse
\citep[for details see][]{SP:10}.
This is then followed by an improvement to
the sky subtraction based on subtracting a combination of principal component
templates \citep{SP:10}. This reduces the amplitude of
the sky subtraction emission line residuals to below 1\% in most cases.

Strong atmospheric telluric absorption features in the red part of the spectrum need to
be corrected for. The telluric correction involves constructing a flux and variance weighted
combination of all the spectra in a given field, which is then iteratively clipped to remove
residual emission or absorption features (such as galaxy emission lines).
This process relies on the fact that in any one field there are a broad range of galaxy redshifts
so features are not present at the same wavelength in many spectra. The resulting average
spectrum is fit by a low order polynomial in the region around the telluric features, while
excluding the regions where the absorption is present. Dividing through by this polynomial
fit results in a telluric correction spectrum which is set to be equal to unity everywhere outside
of the telluric absorption bands. The correction is then implemented by dividing
each individual spectrum by this correction spectrum. 

Finally, 2{\sc dfdr} splices together the blue and red spectra for
each galaxy by doing a first-pass flux calibration
to best match the spectra at the splice-wavelength (5700\,\AA).
The pixel scale is $1\,$\AA\ in the blue and $1.6\,$\AA\ in the red, although during the
splicing step the red spectra are resampled to the same pixel scale as the blue, so the
final pixel scale is $\sim 1\,$\AA\,pix$^{-1}$. This is done with a
quadratic interpolation ensuring that flux is conserved, and with appropriate treatment of
masked or otherwise bad values. The same resampling is
applied to the variance arrays to correctly propagate the errors. The overlap region between
the blue and red spectra is $250\,$\AA, from typically 5650\,\AA\ to 5900\,\AA, although this
varies slightly from spectrum to spectrum depending on the location on the detector.

As 2{\sc dfdr} is continuing to be developed and improved, and to mitigate against
reduction mistakes at the telescope during observing, the entire GAMA dataset is
periodically re-reduced, to ensure that the final spectroscopic data products are
homogeneous and internally self-consistent. The version of 2{\sc dfdr}  used in
producing the final GAMA I spectra was 2{\sc dfdr} v4.42.

Fig.~\ref{fig:spectrum} shows examples of GAMA spectra after the 1D extraction
and flux-calibration process (see \S\,\ref{corrections} below) are completed. This Figure
shows two high quality spectra, and two poor
quality spectra illustrating some of the instrumental and processing
limitations in the survey. The two high quality spectra are an emission line object and
an absorption line object, both with redshift quality of $nQ=4$ \citep[see][for the definition of redshift
quality flags and conventions]{Dri:11}.
The first poor quality spectrum shown gives an example
of fringing, visible as the high-frequency oscillation in the continuum level, and
accompanied by poor removal of the sky features \citep{Sha:06,Sha:13}.
The fringing, which is time-variant, is only present for some fibres, and arises due to air gaps in the
glue between the prism and the ferrule. Over time these fibres have been re-terminated with
new glue and new ferrules. While it doesn't resolve the problem for existing GAMA spectra that are
affected, the AAO has recently completed a total replacement of all 2dF fibres with optimal glue and
ferrules that has now eliminated this problem. In this particular example spectrum a
redshift is still able to be reliably measured, with $nQ=4$.
The second poor quality spectrum shown is an example of a bad splice, characterised
by a dramatic change in continuum level at the splice wavelength of $5700\,$\AA. This feature is a
consequence of poor continuum level estimation due to poor flat-fielding in one or both of the
blue and red arms, in estimating how to scale the two components for the splice. The reliability
of the splicing of AAOmega spectra is an area that is the subject of ongoing work, both at the AAO with
continued development of 2{\sc dfdr}, and within GAMA through investigation of independent flux
calibration processes for the blue and red arms separately. For this spectrum, again a redshift is
still able to be measured with high reliability, $nQ=4$. These examples
illustrate a key point, that a high quality redshift measurement can be obtained from a poor
quality spectrum, although subsequent measurements of emission or absorption features for that
spectrum may not be reliable.

\subsection{Redshift measurements}
\label{redshifting}
Redshifts are measured from the one-dimensional galaxy spectra as soon as each field is
fully reduced using the above process (i.e., typically on the night of observation or the
following day). This is done using the GAMA-specific version
of {\sc runz}, originally developed by Will Sutherland for the 2dFGRS, and now maintained
by Scott Croom. This process is described in \citet{Dri:11}, and further details will be
presented by Liske et al.\ (in prep), including the re-redshifting analysis, which quantifies
the reliability of each measured redshift by having multiple team members re-measure redshifts
for all low-quality flagged measurements, and for a subset of high-quality redshifts.

The {\sc runz} code uses a cross-correlation approach to identifying an automated redshift,
but allows the user to manually identify a redshift in the event of a poor result from the cross-correlation,
before allocated a redshift quality flag ($Q$), from 0 to 4, with 4 being a certain redshift, 3 being
probably correct, 2 indicating a possible redshift needing independent confirmation, 1
indicating that no redshift could be identified, and 0 meaning that the spectrum is somehow flawed
and needs to be reobserved. Following this initial inspection, the process is repeated by multiple
team members, in order to define a robust, probabilistically defined ``normalised" quality scale, $nQ$
\citep{Dri:11}. Details of the error estimates on the redshifts are provided by \citet{Dri:11}. The
re-redshifting process and derivation of the probabilities associated with the $nQ$ quality will be
presented in Liske et al, (in prep.).

A small fraction ($\sim 3\%$) of the GAMA spectra are affected by fringing, as
determined manually during the redshifting and re-redshifting processes. Of these
$\sim 50\%$ still yield a good quality redshift, although other spectroscopic
measurements such as emission line properties (see \S\,\ref{measure} below) are likely to be unreliable.

\begin{figure}
\centerline{\includegraphics[width=80mm]{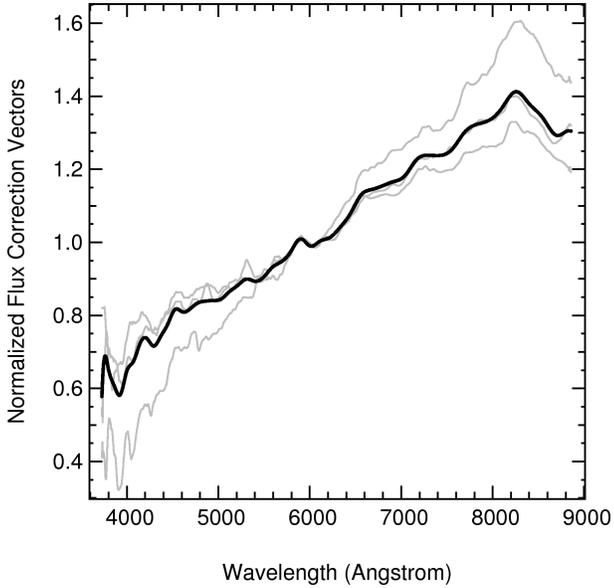}}
\caption{The flux correction vectors from each standard for one 2dF plate,
shown here as an example. The grey lines show the individual flux correction vectors
for each standard star. The black line shows the B-spline fit to the mean vector.
\label{fig:specphoto}}
\end{figure}

\section{Spectrophotometric calibration and quality}
\label{corrections}
\subsection{Flux calibration}
The main purpose of flux calibration is to first correct the wavelength-dependence of the
system throughput (atmosphere, residual wavelength dependence of fibre entrance losses
after the ADC, optics, and CCD quantum efficiency), and second
to provide an approximate absolute flux calibration.
Obtaining accurate spectrophotometry for the GAMA survey presents a challenge due
to the $2''$ optical fibres used for spectroscopy, in addition to observing in conditions that
are not always photometric. Starting with the two-dimensional spectral output from 2{\sc dfdr},
we spectrophotometrically calibrate the data following the {\tt idlspec2d}
pipeline used for the SDSS DR6 \citep{Ade:08}.  We determine a curvature
correction and relative flux calibration for each plate from the standard stars observed on each plate.
The absolute spectrophotometric calibration is determined such that the flux of each object spectrum
integrated over the SDSS filter curve matches the petrosian magnitude of the SDSS photometry for
that object.

\begin{figure}
\centerline{\includegraphics[width=80mm]{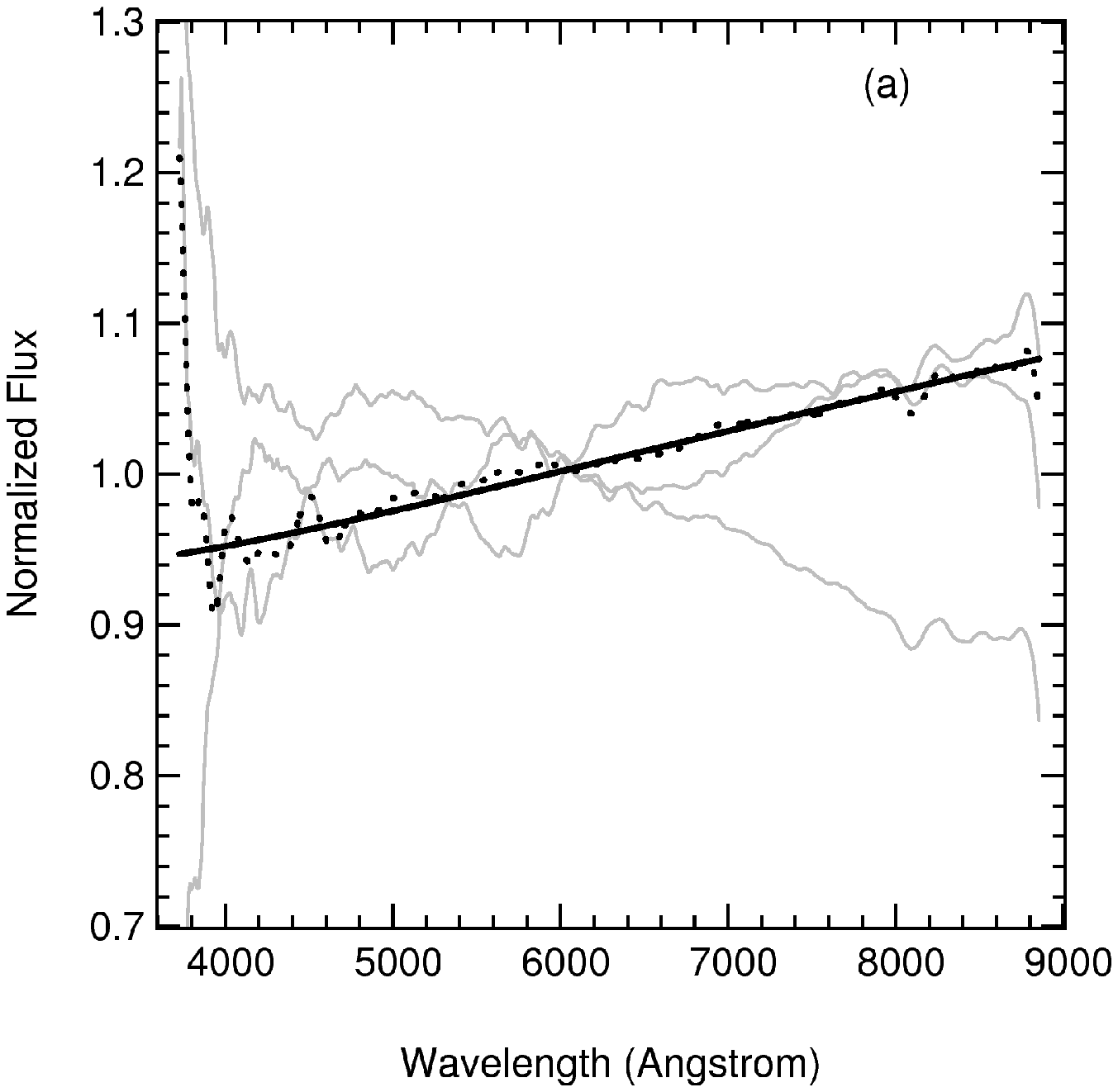}}
\centerline{\includegraphics[width=80mm]{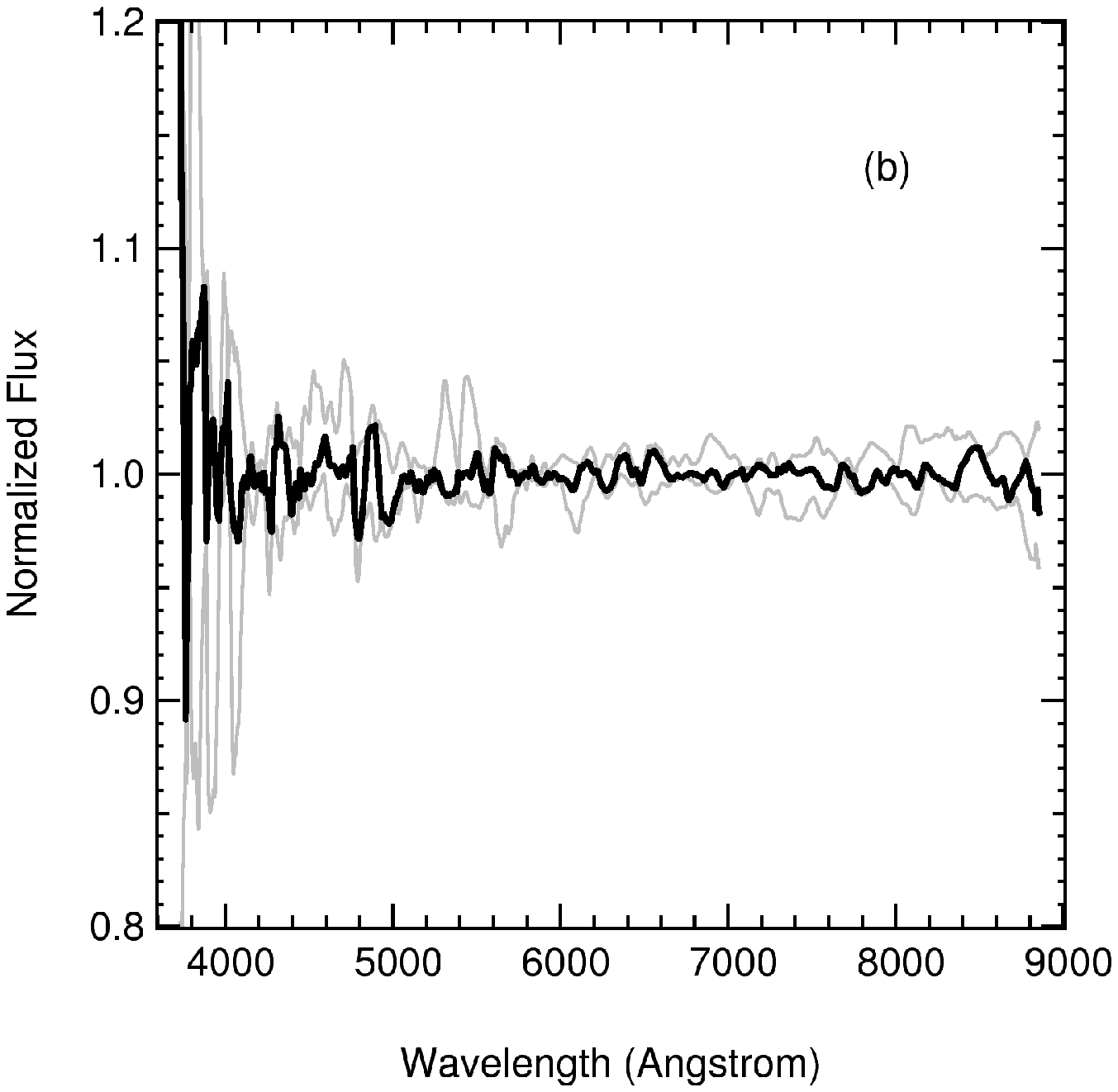}}
\caption{(a) After dividing out the average flux correction vector (Fig.~\ref{fig:specphoto})
to remove the low order shape, the low order residuals for each standard star are shown in grey.
The residuals are then fit with a 4th order Legendre polynomial and the median of these polynomials
is shown by the black line. The dotted line is the B-spline fit.
(b) The high order terms are fit by first dividing out the low order (grey) and then taking the
median of the result (black).
\label{fig:specphoto2}}
\end{figure}

\begin{figure}
\centerline{\includegraphics[width=80mm]{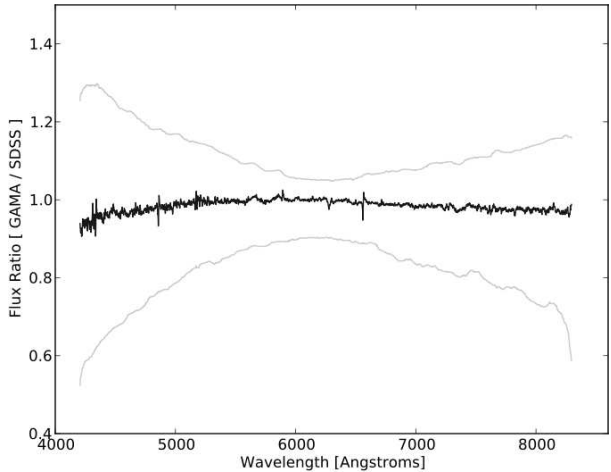}}
\caption{The median ratio of common GAMA and SDSS spectra. This result is derived from 574 objects out of
the 637 objects that were observed by both surveys. The 63 spectra excluded were due to particularly
noisy GAMA spectra, or redshift mismatches between the surveys. The spectra are first normalised by
the median flux value of the flux-calibrated spectrum, (since the absolute flux
calibration is scaled differently between GAMA and SDSS). The spectra are then median filtered by $7\,$\AA\ and
interpolated to the same wavelength scale before the ratio was taken. The dark line shows the
median of the flux ratios, and the outer, grey lines show the 68th percentile range
of the distribution of ratios for individual objects.
\label{fig:sdss_gama_ratio}}
\end{figure}

We typically assign three of the fibres on each 2dF plate for observing standard stars.
The spectroscopic standards are typically colour-selected to be F8 subdwarfs, similar in
spectral type to the SDSS primary standard BD+17 4708.
The spectrum of each standard star is spectrally typed by
comparing it to a set of theoretical spectra generated from Kurucz model atmospheres \citep{Kur:92},
using the spectral synthesis code SPECTRUM \citep{GC:94, Gra:01}. A flux correction vector,
a one-dimensional array of wavelength-dependent
correction factors tied to the wavelength scale,
is derived for each standard star by taking the ratio of its spectrum \citep[in units of counts and after
correcting for Galactic reddening,][]{Schl:98} to its best-fit model (in units of ergs\,s$^{-1}$\,cm$^{-2}$\,\AA$^{-1}$).
This is illustrated in Fig.~\ref{fig:specphoto}.
There are a small number of plates (2 out of a total of 392 for v08 of the GAMA data)
that included no standard star observations. For these plates, we used the standard stars observed on
the plate observed either just before or just after the plate lacking standard stars.

We first calculate an average flux correction vector from the standard stars on each plate.
We select high signal-to-noise regions of the standard star
spectra, and divide out this average correction vector (Fig.~\ref{fig:specphoto2}a).
By removing the overall average of the low-order shape in the standard star continuum in this way, we can fit
the residuals to derive a plate-specific average curvature correction, to account for the declining
CCD response at the extreme blue and red wavelengths. These residuals are fit with 4th order Legendre polynomials (Fig.~\ref{fig:specphoto2}a), and the average low order residual is found by taking the median of the Legendre coefficients. For spectral regions where the standard star spectra have $S/N>12$, higher
order fluctuations are also corrected. These higher order terms are found from the median Legendre
coefficients after the low
order terms are divided out. The flux calibration vector appropriate for each plate is constructed by multiplying the
lower order residual fit and the high order residual fit. This vector is then fit with a B-spline, and the coefficients are
used as the final flux calibration vector.
Once the final flux calibration vector has been determined, it is applied to each individual spectrum
on the plate, resulting in each spectrum being correctly flux calibrated, in a relative sense.

The final step in obtaining an absolute flux calibration then
involves tying the spectrophotometry directly to the $r$-band Petrosian
magnitudes measured by the SDSS photometry. This is accomplished by multiplying each individual
spectrum by the SDSS $r$-band filter response. The SDSS magnitudes are based on photon counting,
so this calculation is done by integrating $f_{\nu}/{\nu}$ times the filter transmission function.
We then determine the ratio of this flux with that
corresponding to the SDSS Petrosian $r$-band magnitude for each object.
The GAMA spectra are then linearly scaled
according to this value. It is straightforward to modify this scaling factor if alternative photometric references
are preferred. In particular, we are exploring the utility of Sersic $r$-band magnitudes
as an alternative, although for the present the direct comparisons to SDSS measurements
provide important consistency checks.

To test how well the applied flux calibration method agrees with the
methods applied to SDSS fibre spectroscopy, we look to objects observed by both surveys.
There are $637$ objects observed by both SDSS and GAMA, which are mostly galaxies,
but also including $120$ standard stars. After removing a small number of spectra to exclude particularly
noisy spectra or those with mismatched redshifts, Fig.~\ref{fig:sdss_gama_ratio} shows the median and 68th
percentiles of the ratio of the normalised GAMA and SDSS spectra as a function of wavelength, after the flux
calibration is applied to the GAMA spectra. The spectra are normalised by the median flux of each spectrum
before taking the ratio, due to the different absolute flux calibrations applied (to the fibre magnitude in SDSS,
and to the Petrosian magnitude in GAMA). It is clear from the solid line that the
flux calibration applied to the GAMA spectra results in a good agreement
across the entire wavelength regime with high fidelity, and best in the mid-wavelength range of the spectra.
The extreme blue and red ends of the spectrum are noisier, and worst in the blue, but the overall agreement
resulting from the flux calibration applied to the AAOmega spectra from GAMA is still robust.
Overall, we find an accuracy of around 10\% in the flux calibrated spectra, ranging to somewhat worse than 20\%
at the extreme wavelength ends of the spectra.
There remains an unresolved issue associated with the level of the
response in the blue end of the spectra, which is the focus of ongoing work. For this reason, we
advise caution when working with the bluest spectral diagnostics (such as the [OII] emission line
and the $4000\,$\AA\ break) in the current generation of measurements.
Based on this analysis, we estimate that the flux calibrated GAMA
spectra are typically accurate to between $\sim 10-20\%$, although the small fraction of poorly spliced
spectra and those affected by fringing are likely to be much worse. We note that this precision has been
estimated using bright spectroscopic targets. For the fainter GAMA spectra it is likely that the
spectrophotometric precision will not achieve this level, and we are also continuing to work
on quantifying the dependence with magnitude.

\begin{figure*}
\centerline{\includegraphics[width=80mm]{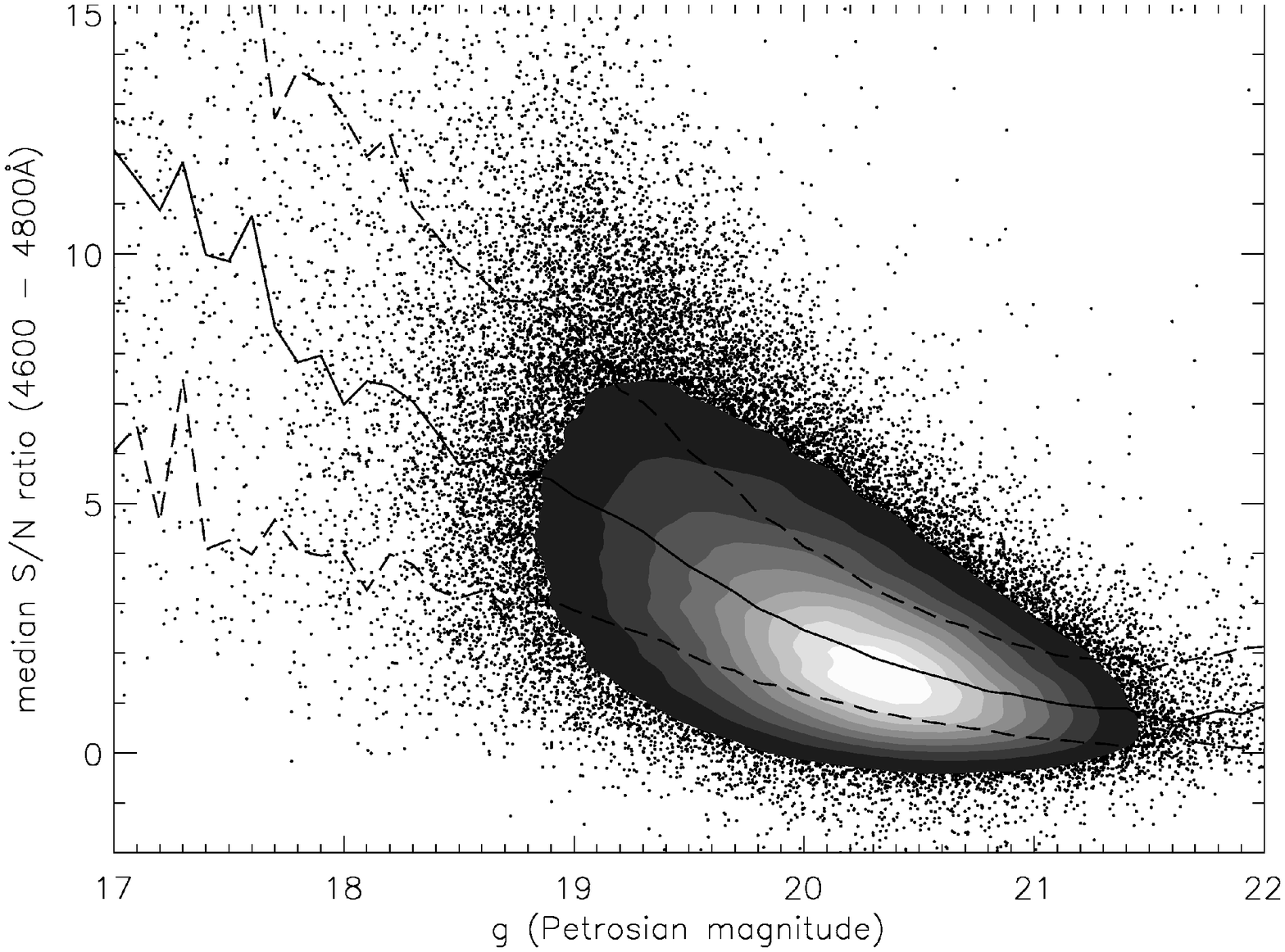}\hfill
\includegraphics[width=80mm]{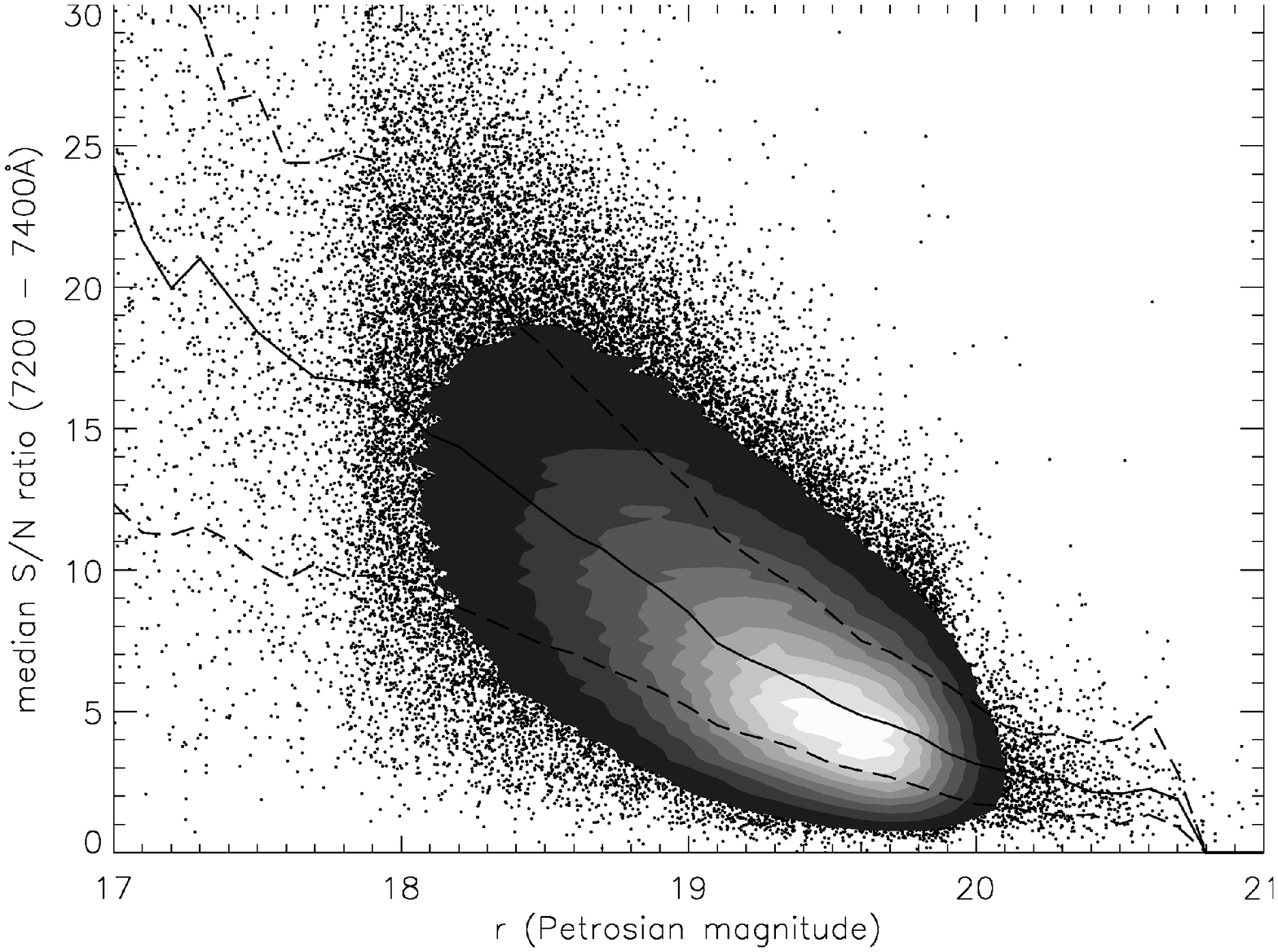}}
\caption{The $S/N$ in the blue (left) and red (right) arms of the GAMA spectra,
as a function of $g$- or $r$-band Petrosian magnitude, respectively. The solid and dashed lines
show the median and 68th percentiles. It is clear that brighter
targets have higher $S/N$, and also that the red arm of the AAOmega spectrograph is
relatively more sensitive for a given exposure time. The magnitude limit of $r\approx17.7$\,mag at
the bright end corresponds to the SDSS main galaxy sample limit of $r=17.77$\,mag.
The GAMA selection limit of $r=19.8$\,mag at the faint end is also apparent. The objects outside
these limits include bright targets not observed by SDSS due to fibre-collision limitations, and faint
targets included through the additional $K$ and $z$-band selection limits \citep{Bal:10}, together
with specific additional filler targets (Ching et al.\ in prep.).
\label{fig:snmag}}
\end{figure*}

\begin{table}
\begin{center}
\caption{GAMA spectral emission lines measured by GANDALF.}
\label{tab:lines}
\begin{tabular}{cc}
\hline
Emission line & wavelength (\AA) \\ \hline
HeII & 3203 \\
$[$NeV$]$ & 3345 \\
$[$NeV$]$ & 3425 \\
$[$OII$]$ & 3727 \\
$[$NeIII$]$ & 3869 \\
H5 & 3889 \\
$[$NeIII$]$ & 3967 \\
H$\epsilon$ & 3970 \\
H$\delta$ & 4102 \\
H$\gamma$ & 4340 \\
$[$OIII$]$ & 4363 \\
HeII & 4686 \\
$[$ArIV$]$ & 4711 \\
$[$ArIV$]$ & 4740 \\
H$\beta$ & 4861 \\
$[$OIII$]$ & 4959 \\
$[$OIII$]$ & 5007 \\
$[$NI$]$ & 5199 \\
HeI & 5876 \\
$[$OI$]$ & 6300 \\
$[$OI$]$ & 6364 \\
$[$NII$]$ & 6548 \\
H$\alpha$ & 6563 \\
$[$NII$]$ & 6583 \\
$[$SII$]$ & 6716 \\
$[$SII$]$ & 6731 \\
\hline
\end{tabular}
\end{center}
\end{table}

\subsection{Spectral signal-to-noise}
Here we detail the continuum signal-to-noise ($S/N$) distribution for the GAMA spectra as a function of
apparent magnitude (Fig.~\ref{fig:snmag}). The $S/N$ per pixel is quantified by measuring the
median of the ratio of the observed flux to the noise in a $200\,$\AA\ window at the centre of the
blue and red arms. The noise is determined from the variance arrays
described in \S\,\ref{1Dspec}.
In the blue, the wavelength range $4600-4800\,$\AA\ was selected,
and $7200-7400\,$\AA\ in the red. It can be seen that the $S/N$ properties vary as expected, with
brighter targets showing higher $S/N$. The red arm of the GAMA
spectra displays a typical $S/N$ of a few at the faintest observed magnitudes, increasing to well over 10 at the
brightest. In the blue arm the $S/N$ is typically between $1-5$ from the faintest to brightest sources. There are
a very small fraction of spectra where the measured $S/N$ in the blue is negative. This is a consequence
of the continuum measurement being negative for these spectra, and is an artifact arising from
poor flat-fielding or scattered light subtraction. There are also targets shown that extend fainter than our nominal
survey selection limit of $r=19.8$\,mag. These enter the survey as a result of our supplementary
$K$- and $z$-band selection limits \citep{Bal:10}, as well as the addition of a selection of ``filler" targets.
Due to galaxy clustering and the limitation of how closely fibres can be positioned with 2dF,
there are frequently a number of fibres on any given plate that cannot be assigned to a primary
survey target. In this case, and in order to maximise the scientific return from the survey, we
allow fibres to be assigned to supplementary targets selected to support specific scientific programs \citep{Dri:11}.
In particular, we allocate such filler fibres to target systems identified as interesting objects in the
far-infrared by the {\em Herschel}-ATLAS survey, as well as to radio sources with carefully identified optical
counterparts, that otherwise lack redshift or spectroscopic measurements (Ching et al.\ in prep.).

\begin{figure*}
\centerline{\includegraphics[width=75mm]{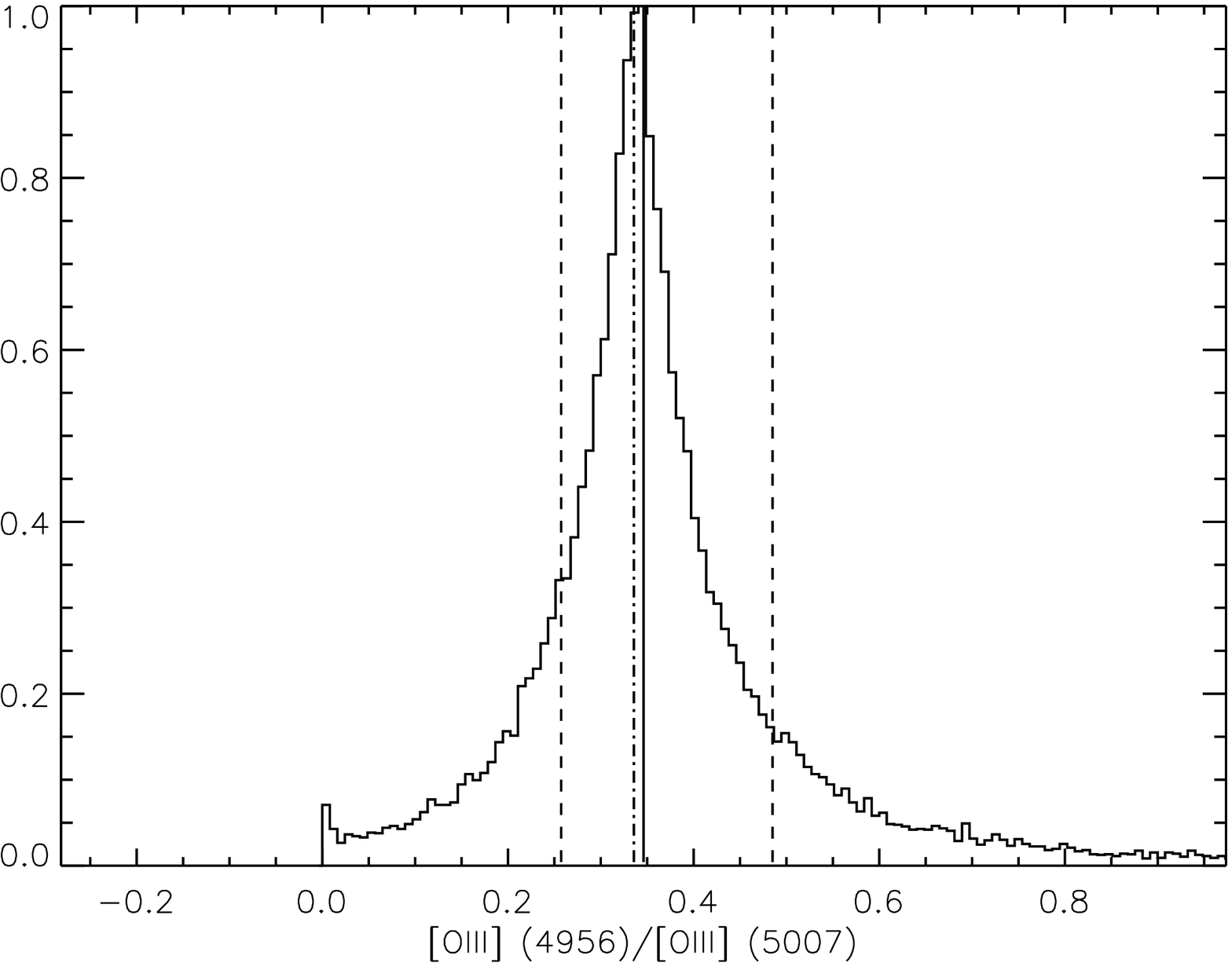}\hfill
\includegraphics[width=80mm]{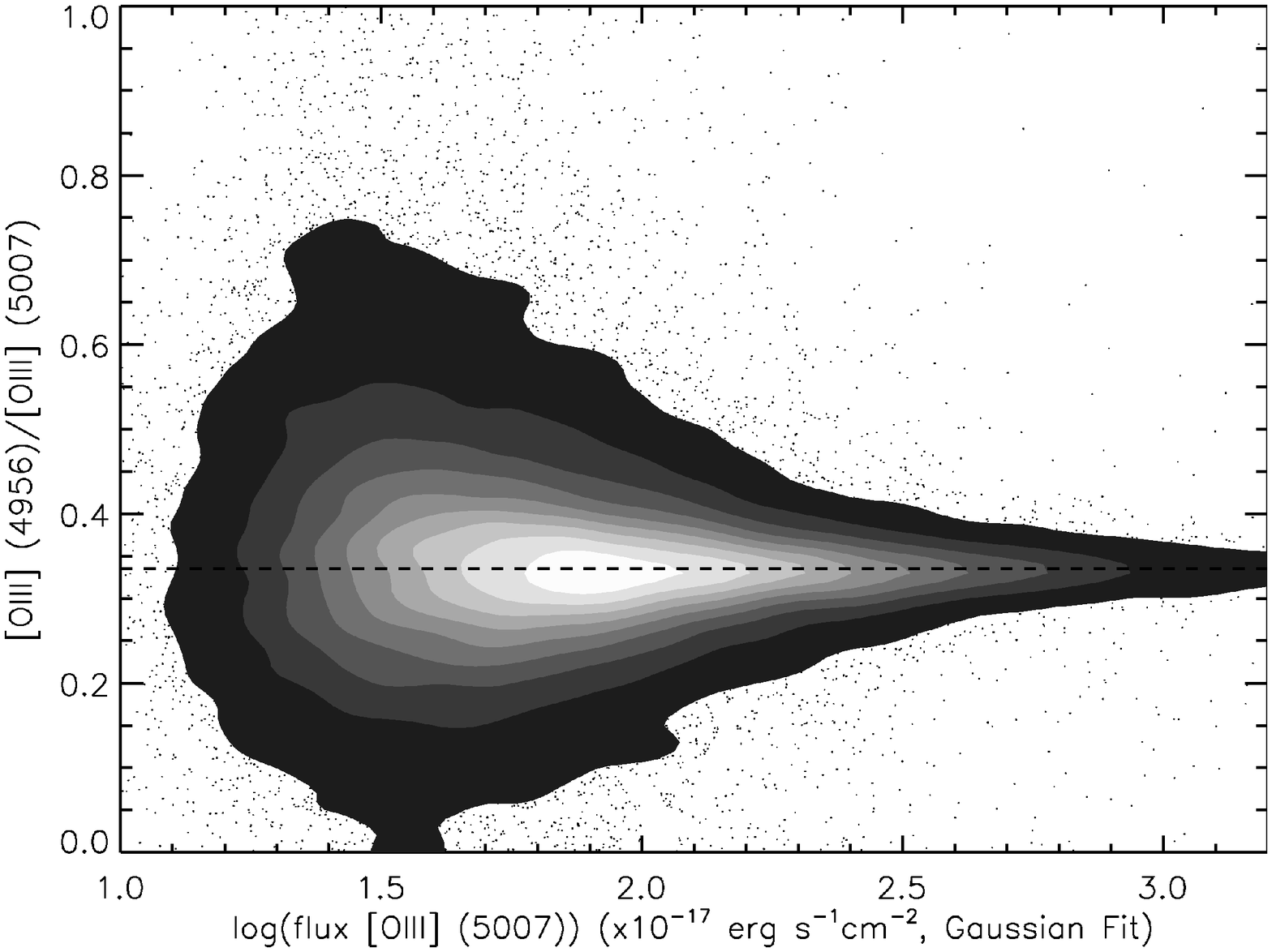}}\vspace{2mm}
\centerline{\includegraphics[width=75mm]{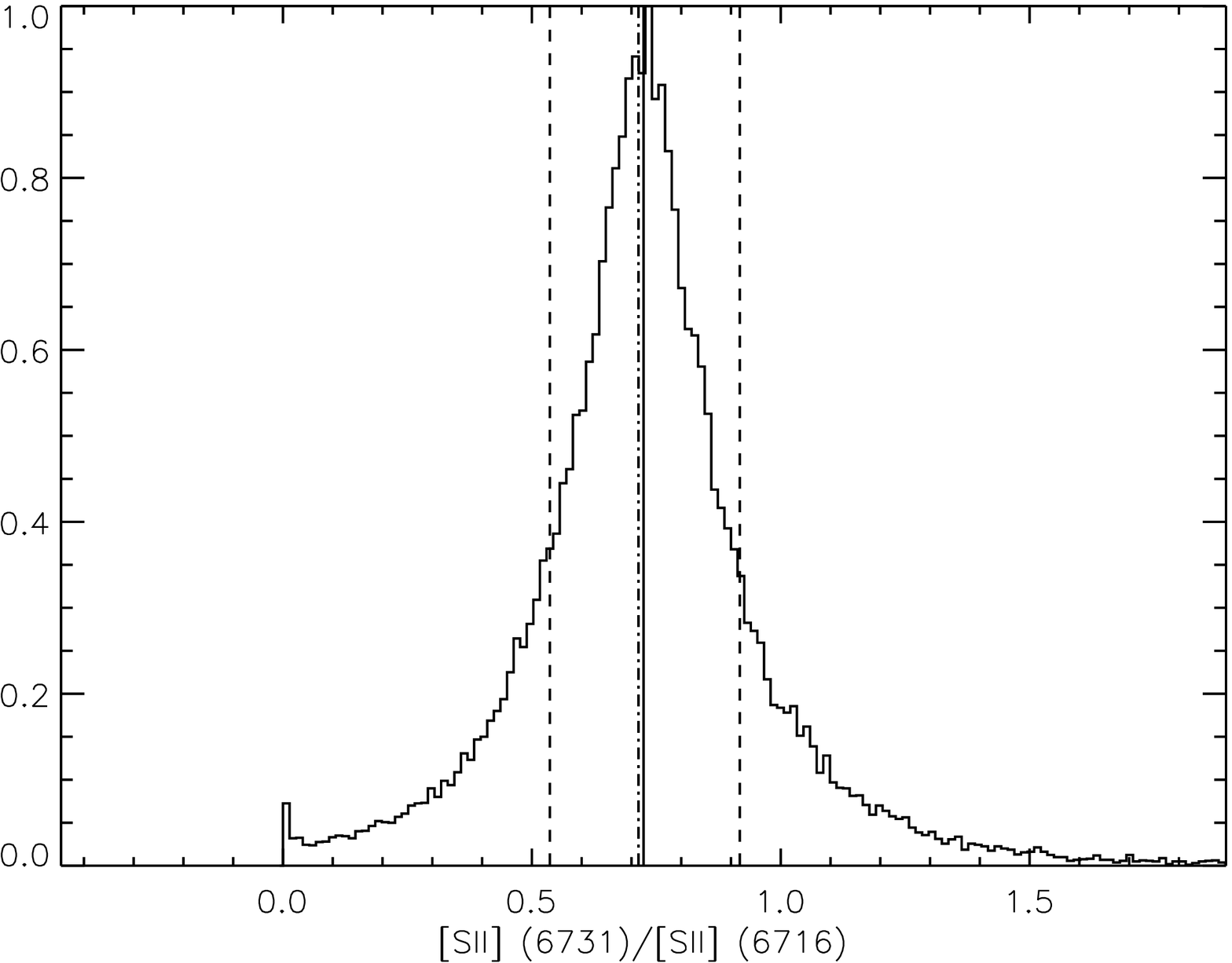}\hfill
\includegraphics[width=80mm]{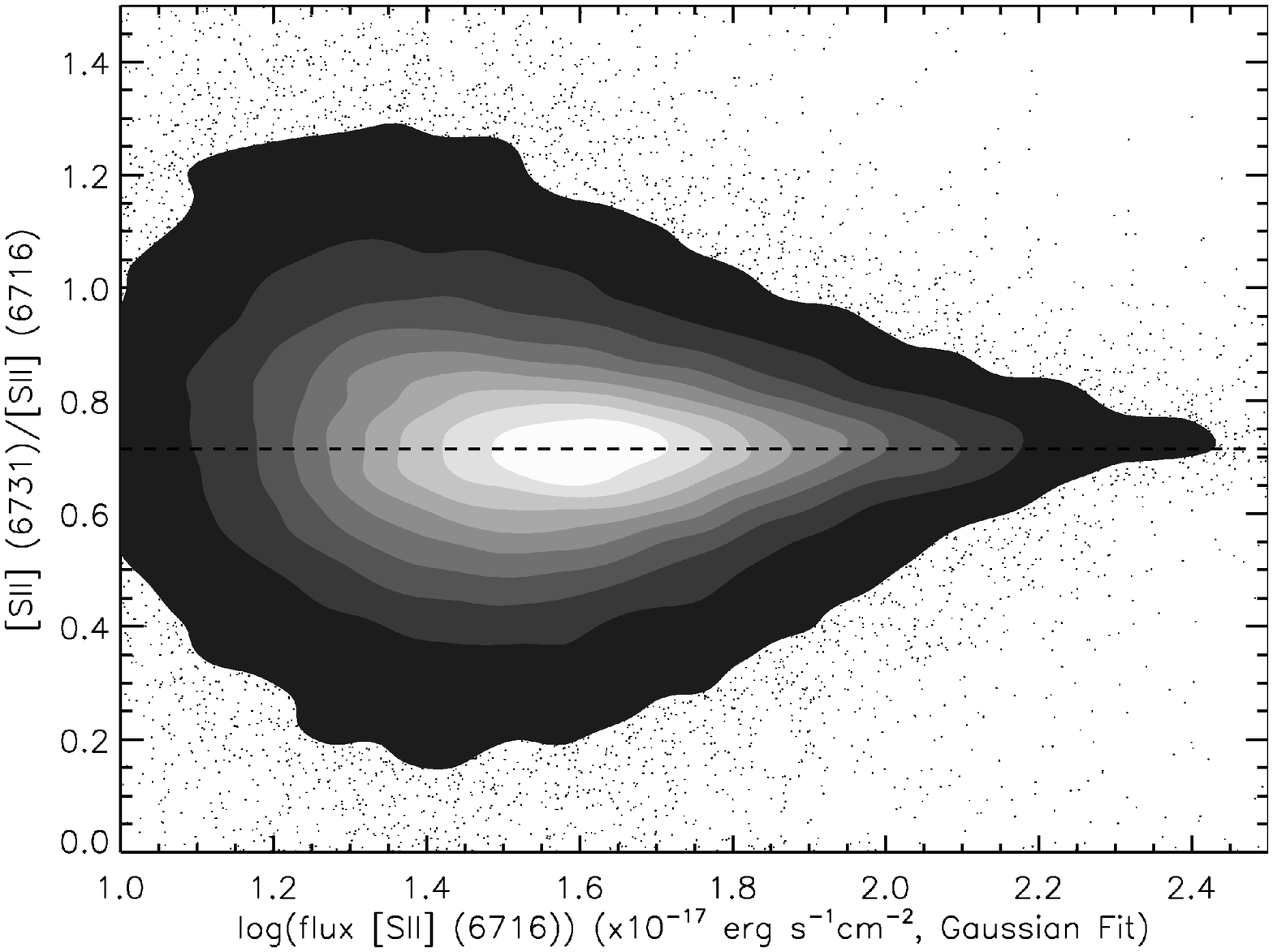}}
\caption{Upper left: The distribution of the ratio of [OIII]$\lambda4956$/[OIII]$\lambda5007$, showing consistency
with the value fixed, for a given density and temperature, by quantum mechanics. The solid line is the median,
and the dashed lines are the
68th percentiles. The expected ratio of $1/2.98$ \citep{SZ:00} is shown as the
dot-dashed line. Upper right: The flux ratio shown now as a function of the flux in the brighter line. The dashed
line here is the expected value. The bottom panels reproduce this analysis for
[SII]$\lambda6716$/[SII]$\lambda6731$, where the expected ratio is $1/1.4$ \citep{Ost:89}.
\label{fig:lineratios}}
\end{figure*}

\section{Measurement of spectral features}
\label{measure}
The flux-calibrated 1D spectra are those used for all subsequent measured spectroscopic properties.
Both emission lines and absorption features are measured. A variety of standard galaxy properties
are catalogued based on these measurements, including Balmer decrements
and star formation rates from the H$\alpha$ and H$\beta$ emission lines \citep{Gun:13},
spectral diagnostics \citep{BPT:81} to discriminate AGN from star forming objects \citep{Gun:13},
metallicity measurements for the nebular gas from the emission line features \citep[Lara-L{\'o}pez et al., in prep.]{Fos:12},
stellar velocity dispersions from absorption features and the D4000 age-estimate
parameter from the 4000\,\AA\ break, and more. This section details the
measurement processes for quantifying the spectral features used in deriving these now
standard properties for GAMA galaxies. The measurement of the specific properties themselves
are detailed in the various papers presenting the analysis of those parameters, referenced above.

Emission lines are measured in two ways. We first fit Gaussians to a selection of common
emission lines at appropriate observed wavelengths, given the measured redshift of
each object. This is performed for a subset of common emission lines
via a simultaneous iterative $\chi^{2}$ fitting of positive emission peaks to these
emission lines in three independent line groups, around the wavelengths of
the [OII], H$\beta$, and H$\alpha$ lines. The local continuum spanning each
fitting region is approximated with a linear fit. All lines in each group are fitted simultaneously.
As well as the continuum coefficients and peak intensities, a small velocity offset (from the
underlying GAMA redshift) is allowed, and a line width common to all lines in the group
is also fit. Line width is constrained to lie between 0.5-5 times the instrumental
PSF (3.4 CCD pixels, 3-5\AA). Each line group is fitted with independent values for the free
parameters to accommodate small local variations in the spectral sampling. Flux values for
individual lines are rejected if inclusion of a line in the global fit fails to improve the reduced
$\chi^2$ value significantly (by a factor of 3). Typical GAMA sources exhibit only marginally
resolved emission lines at the resolution of the AAOmega spectra. No attempt has been made to
fit multiple emission components to composite emission line structures, such as the [OII] doublet,
at this time. Integrated line flux, EW and the associated statistical errors are estimated from the fitting process.
The errors are those associated with the formal Gaussian fitting process, and as
shown in \S\,\ref{dupobs} below, the error estimates are robust although likely to be somewhat
underestimated for fainter galaxies. In addition, flags are provided to identify a variety of problematic
cases, including failed measurements or lines falling in masked spectral regions, lines that are apparently
too narrow to be fit (usually caused by bad pixels), and lines that are fit by the maximum allowed width
of the fitting routine (a Gaussian $\sigma=10\,$\AA), typically due to the presence of an intrinsically broad line.
This provides a baseline set of emission line fluxes, equivalent widths, and $S/N$ estimates. 
For those (small fraction of) objects such as broad-line active galactic nuclei (AGN) and sources
that exhibit line splitting (indicative of either merging systems or starburst winds),
the simple single Gaussian emission approximation will not provide useful estimates.

Independently, we also pursue a more sophisticated spectral measurement approach,
in order to self-consistently derive both stellar kinematics and emission line properties.
To do this, we use the publicly available codes pPXF \citep{Cap:04} and GANDALF \citep{Sar:06}.
We first extract the stellar kinematics using pPXF, matching the observed spectra to a set of stellar
population templates from \citet{Mar:11} based on the MILES stellar library \citep{SB:06},
masking the regions which could potentially be affected by nebular emission, given the observed redshift.
For this step, we downgrade the spectral resolution of the \citet{Mar:11} models to match the GAMA spectral
resolution, adopting a value of $2.54\,$\AA\ (FWHM) for the MILES resolution 
\citep{Bei:11}. Next, we use GANDALF v1.5 to simultaneously fit both Gaussian emission line templates
and the stellar population templates, which are broadened to account for the derived stellar kinematics,
to the data, while also correcting for diffuse (stellar continuum) obscuration.
It is important to note that diffuse obscuration, i.e.\ that caused by diffuse dust in the
galaxy, affects the entire spectrum, and no obscuration correction is applied purely to
emission lines during the fitting process (even though this option is available).
We made this decision in order to minimise the number of poor spectral fits
caused by low spectral $S/N$ that otherwise led to ambiguous emission-line fluxes,
when noise was a more significant factor than reddening.
A \citet{Cal:01} obscuration curve is used in estimating the obscuration
corrections to the stellar continuum, and is applied throughout the GANDALF measurement
process. The nebular emission lines from GANDALF are subsequently corrected for
obscuration using a Milky Way obscuration curve \citep{Car:89}, as recommended by \citet{Cal:01}.

GANDALF's simultaneous fitting mechanism allows us to
accurately extract ionised gas emission from the stellar continuum, minimising contamination
from stellar absorption, in order to calculate emission line fluxes and gas kinematics
(velocities and velocity dispersions) from the Gaussian emission line templates.
The outputs from this analysis are line fluxes and equivalent widths,
velocity dispersions (from the stellar absorption lines in the best-fitting SEDs),
and associated derived products such as Balmer decrement. A list of emission
lines for which flux and equivalent width measurements are extracted is given in
Table~\ref{tab:lines}. The wavelengths given for these lines are the wavelengths as measured in air,
as opposed to the vaccuum wavelengths used by SDSS \citep{Yor:00}.

Our final set of measurements includes the flux, equivalent width and signal-to-noise ratio for
each emission line, as well as a velocity dispersion inferred from the line width.
We calculate an emission line ratio diagnostic classification \citep{BPT:81} for each galaxy
based on these measurements. The stellar velocity dispersion and $E(B-V)$ values from
the SED fits, for both diffuse and nebular obscuration (if applicable), are also recorded.
While stellar velocity dispersions are measured for all spectra, these are typically only
robust for spectra having high $S/N$ \citep[e.g.,][Thomas et al., in prep.]{Pro:08,Shu:11}.
In addition, the best-fit SED template and associated $\chi^2$,
along with a clean emission-line free absorption spectrum are available.

\section{Quantifying measurement reliability}
\label{qc}
We explore the quality of our spectroscopic measurements thoroughly. Here we detail
the internal reliability for measurements within each spectrum (\S\,\ref{intcon}),
the repeatability of our measurements using duplicate observations (\S\,\ref{dupobs}), and the self-consistency of
our measurements between the two independent approaches we use (\S\,\ref{selfcon}). We go on to outline
the impact of stellar absorption on the Gaussian fit Balmer line measurements (\S\,\ref{stelabs}),
the reliability of velocity dispersion estimates (\S\,\ref{vdisp}), and the extent of aperture effects (\S\,\ref{apeff}).

\begin{figure*}
\centerline{\includegraphics[width=60mm]{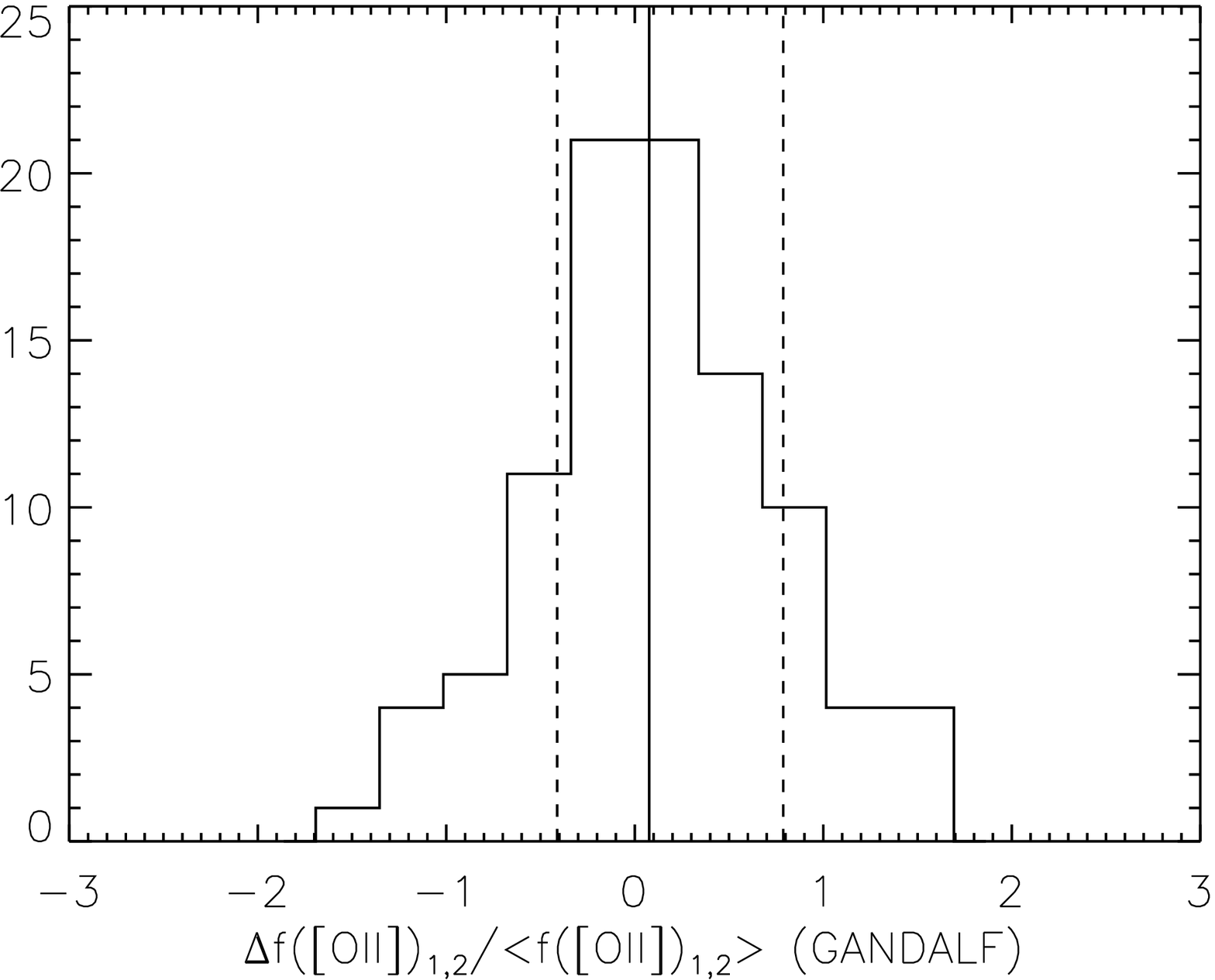}\hspace{1cm}
\includegraphics[width=60mm]{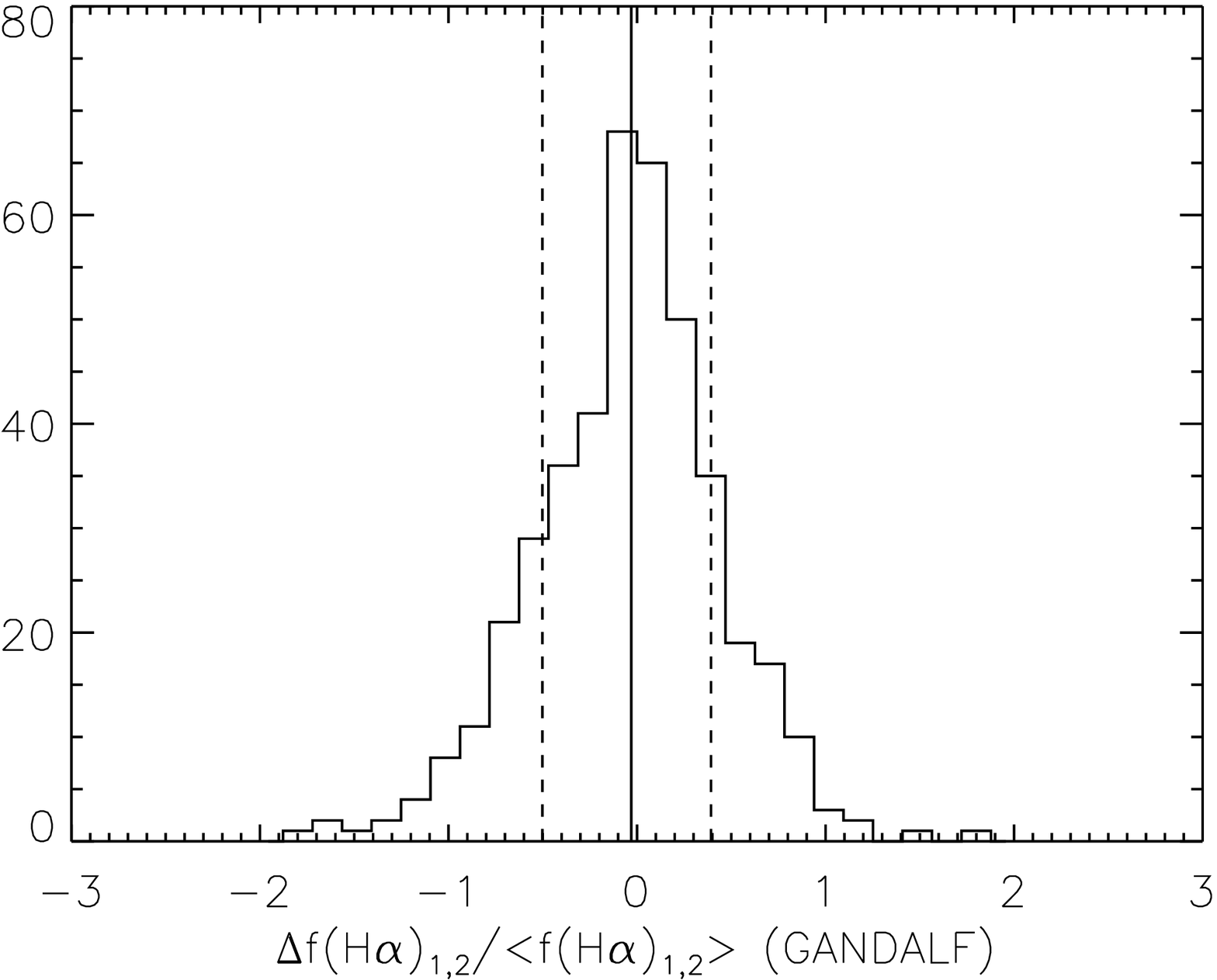}}\vspace{2mm}
\centerline{\includegraphics[width=60mm]{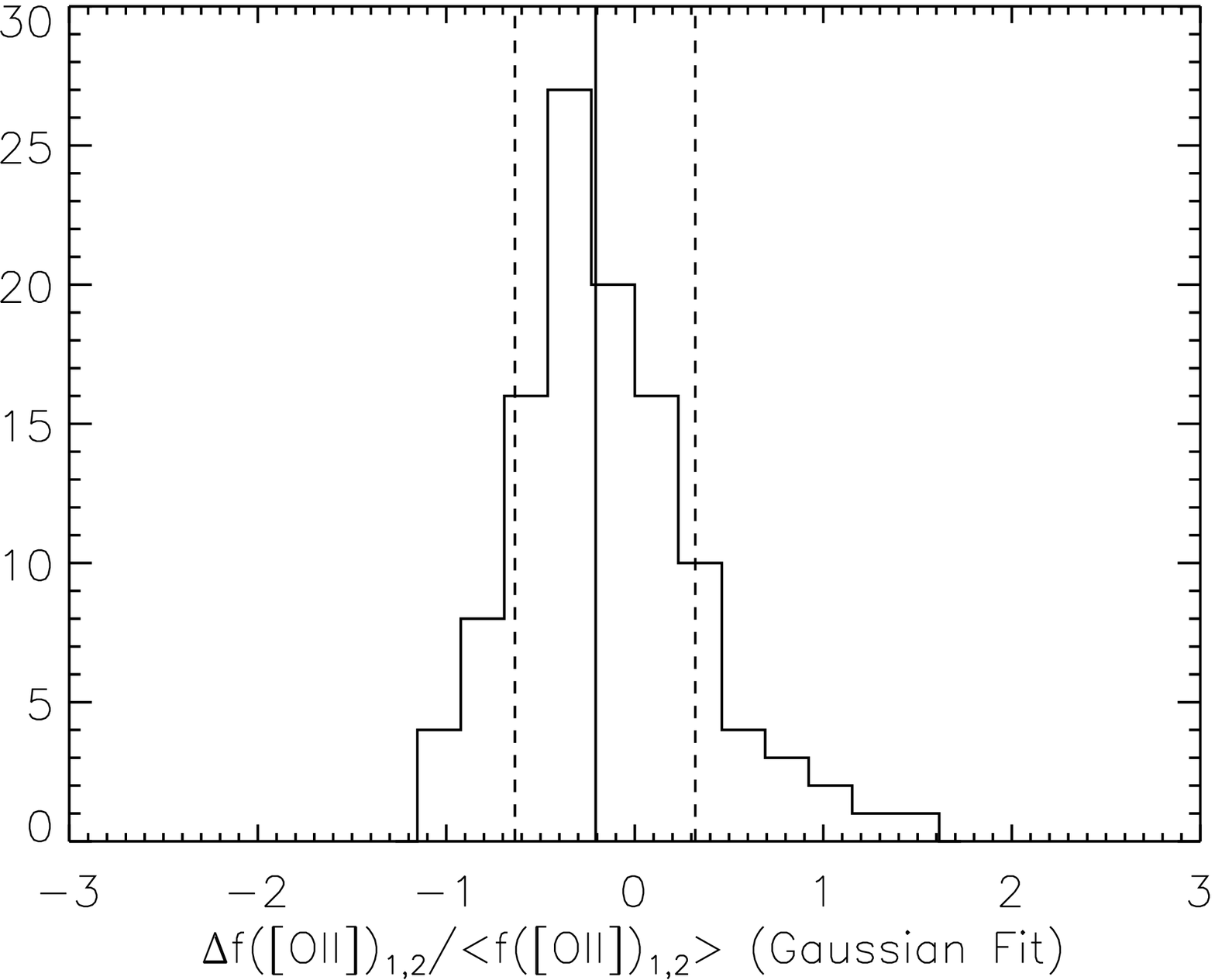}\hspace{1cm}
\includegraphics[width=60mm]{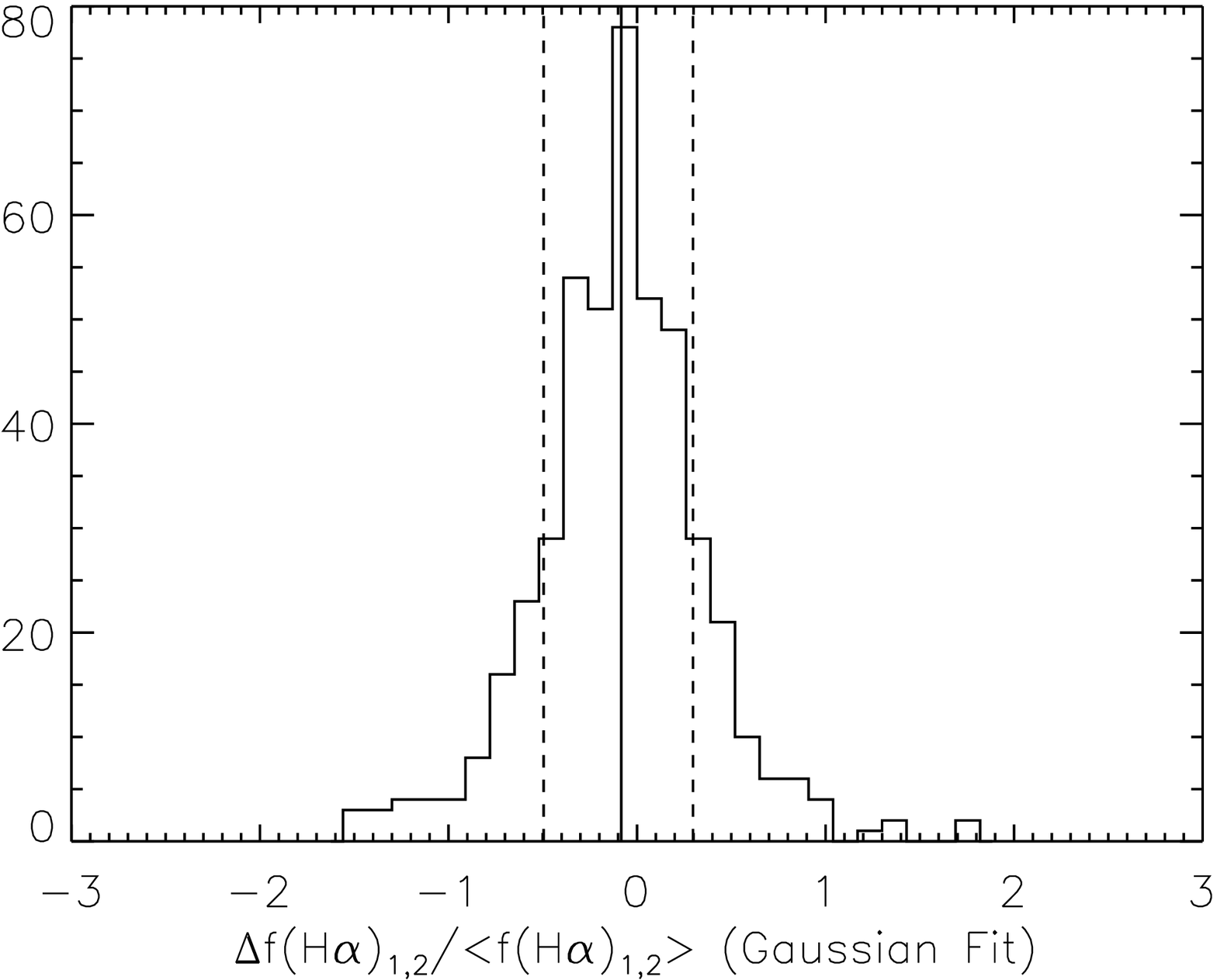}}
\caption{Top row: The distribution of the ratio of the differences in the GANDALF emission line fluxes to the
mean of the two flux measures for [OII] (left) and ${\rm H}\alpha$ (right), for repeat measurements of the same
galaxy. The difference $\Delta f = f_1-f_2$ is taken in the sense that
$f_1$ comes from the spectrum with the higher $S/N$ of the pair of flux measurements. Bottom row: Distribution of
the relative differences for Gaussian fit measurements. In each panel the solid line shows the median of
the distribution, and the dashed lines show the 68th percentiles.
\label{fig:F_dup_comp}}
\end{figure*}

\begin{figure*}
\centerline{\includegraphics[width=60mm]{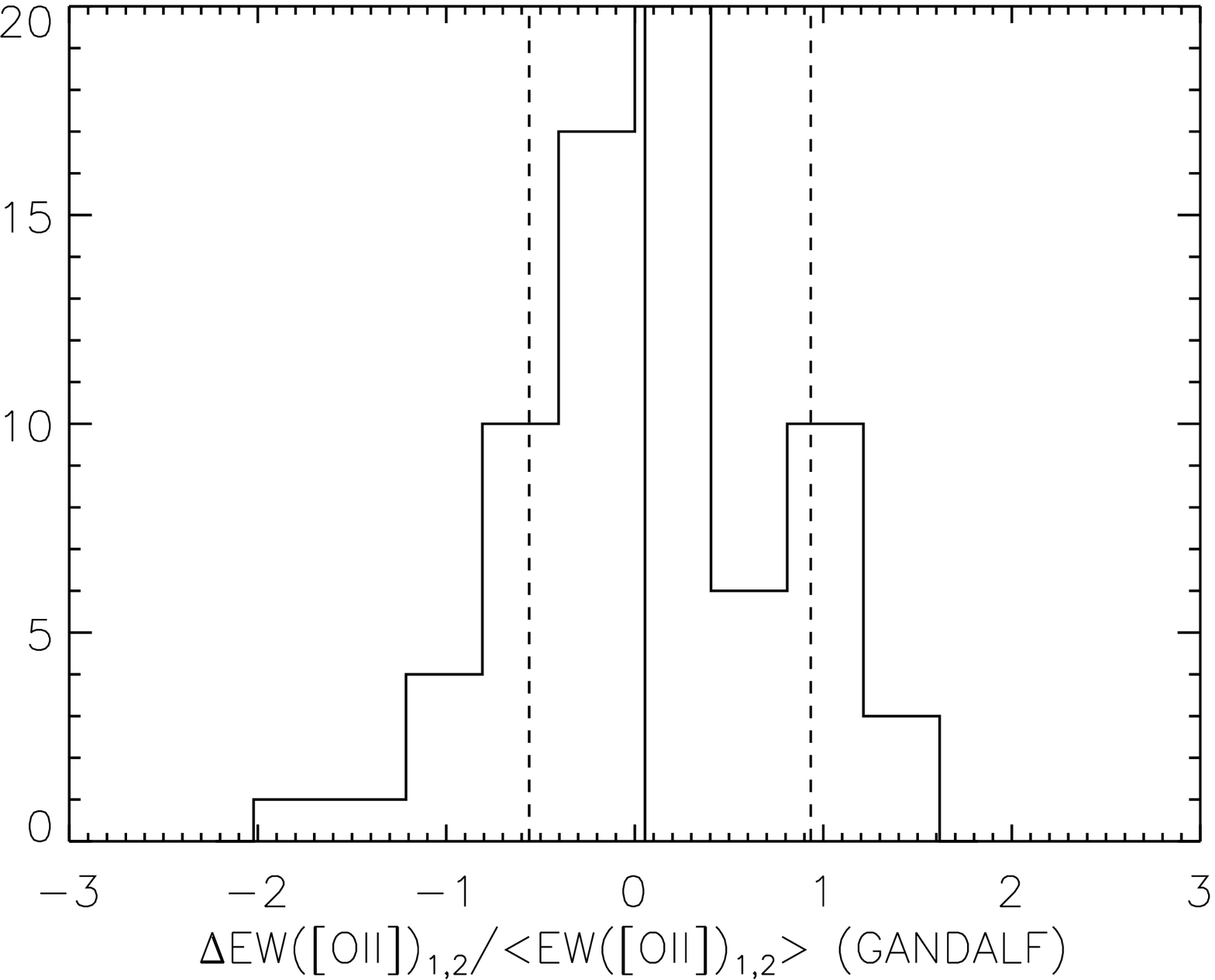}\hspace{1cm}
\includegraphics[width=60mm]{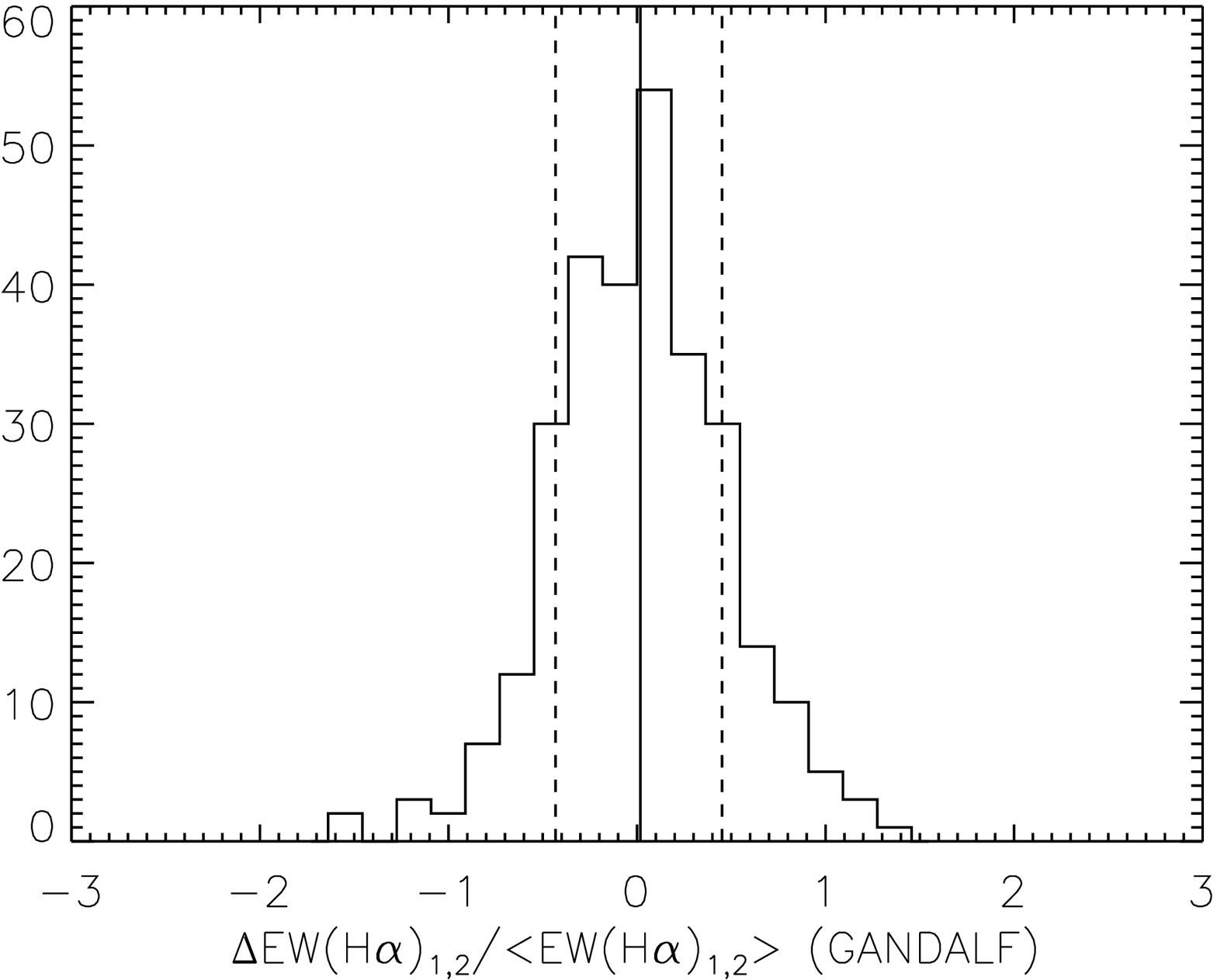}}\vspace{2mm}
\centerline{\includegraphics[width=60mm]{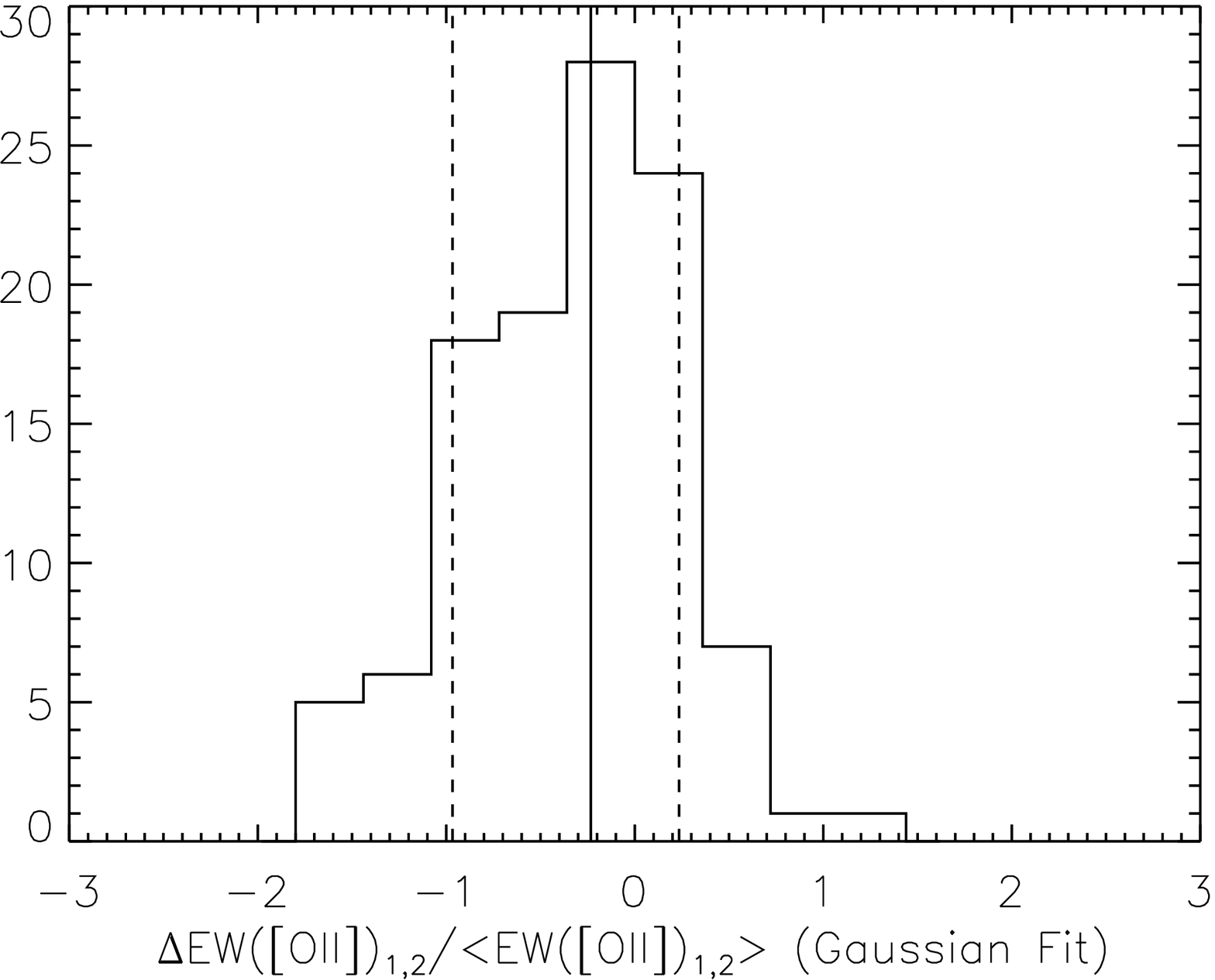}\hspace{1cm}
\includegraphics[width=60mm]{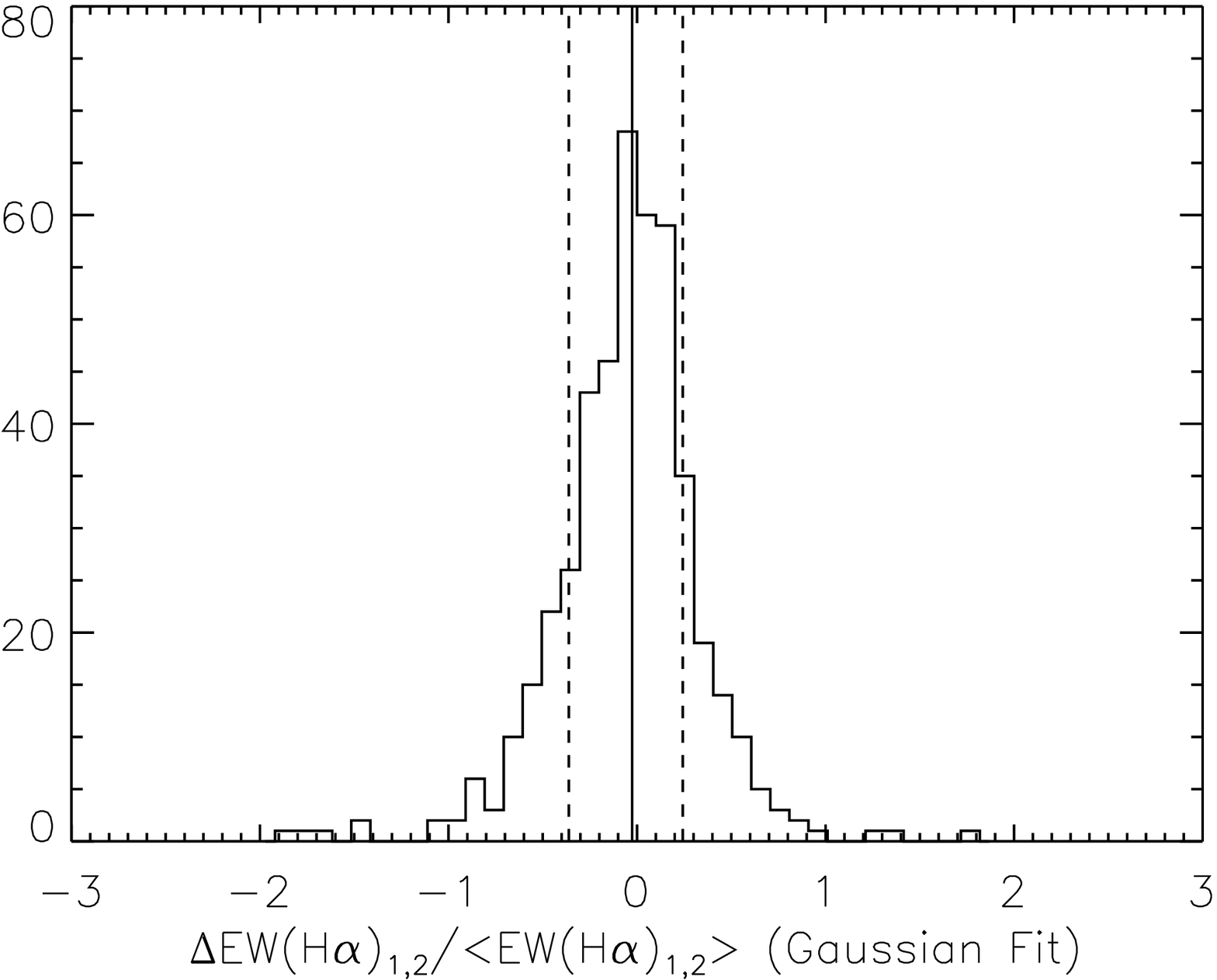}}
\caption{Same as Fig.~\ref{fig:F_dup_comp} but for [OII] (left) and ${\rm H}\alpha$ (right) equivalent widths
rather than fluxes.
The top row shows the GANDALF results, while the bottom row shows the Gaussian Fit results.
\label{fig:ew_dup_comp}}
\end{figure*}

\begin{figure*}
\centerline{\includegraphics[width=60mm]{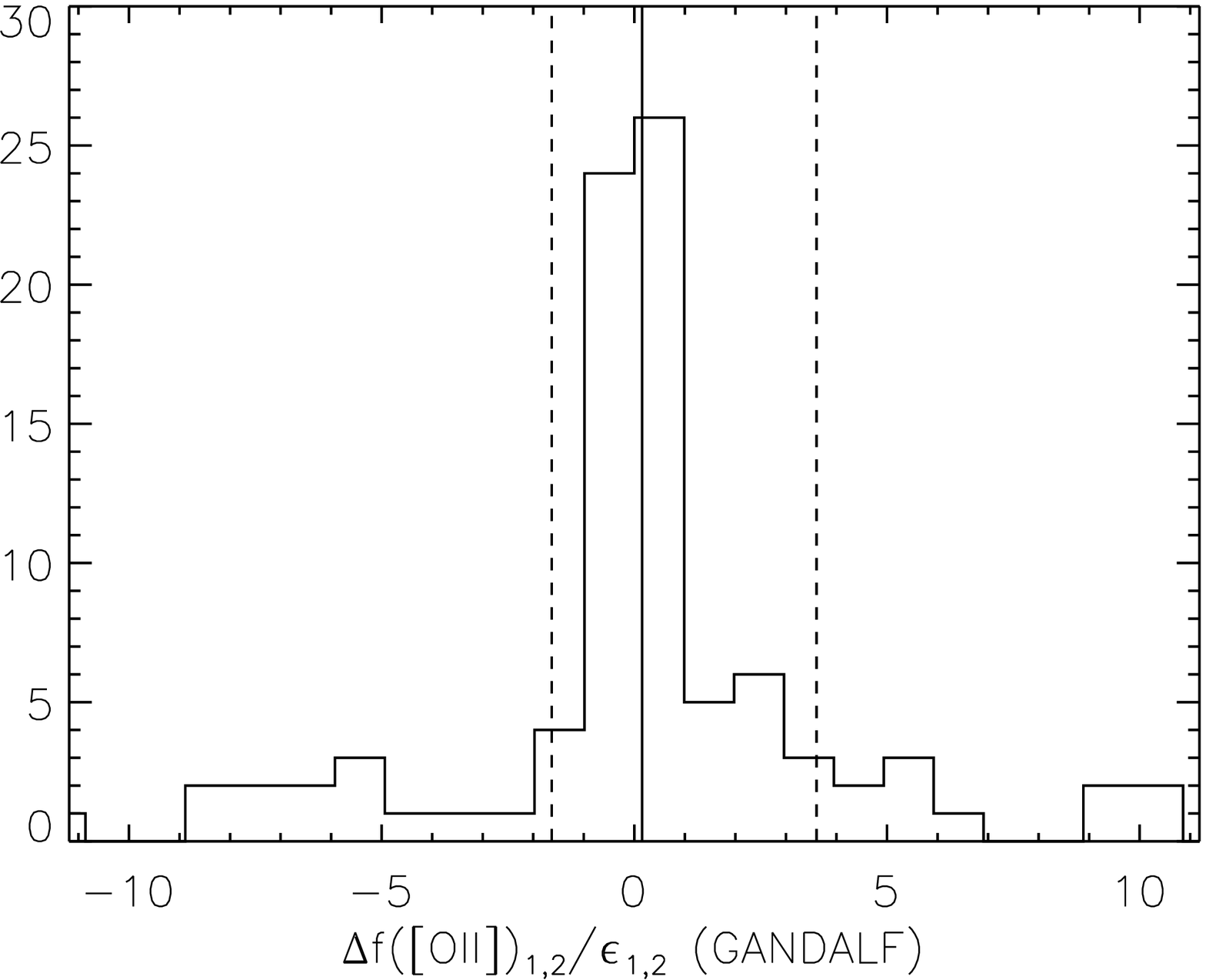}\hspace{1cm}
\includegraphics[width=60mm]{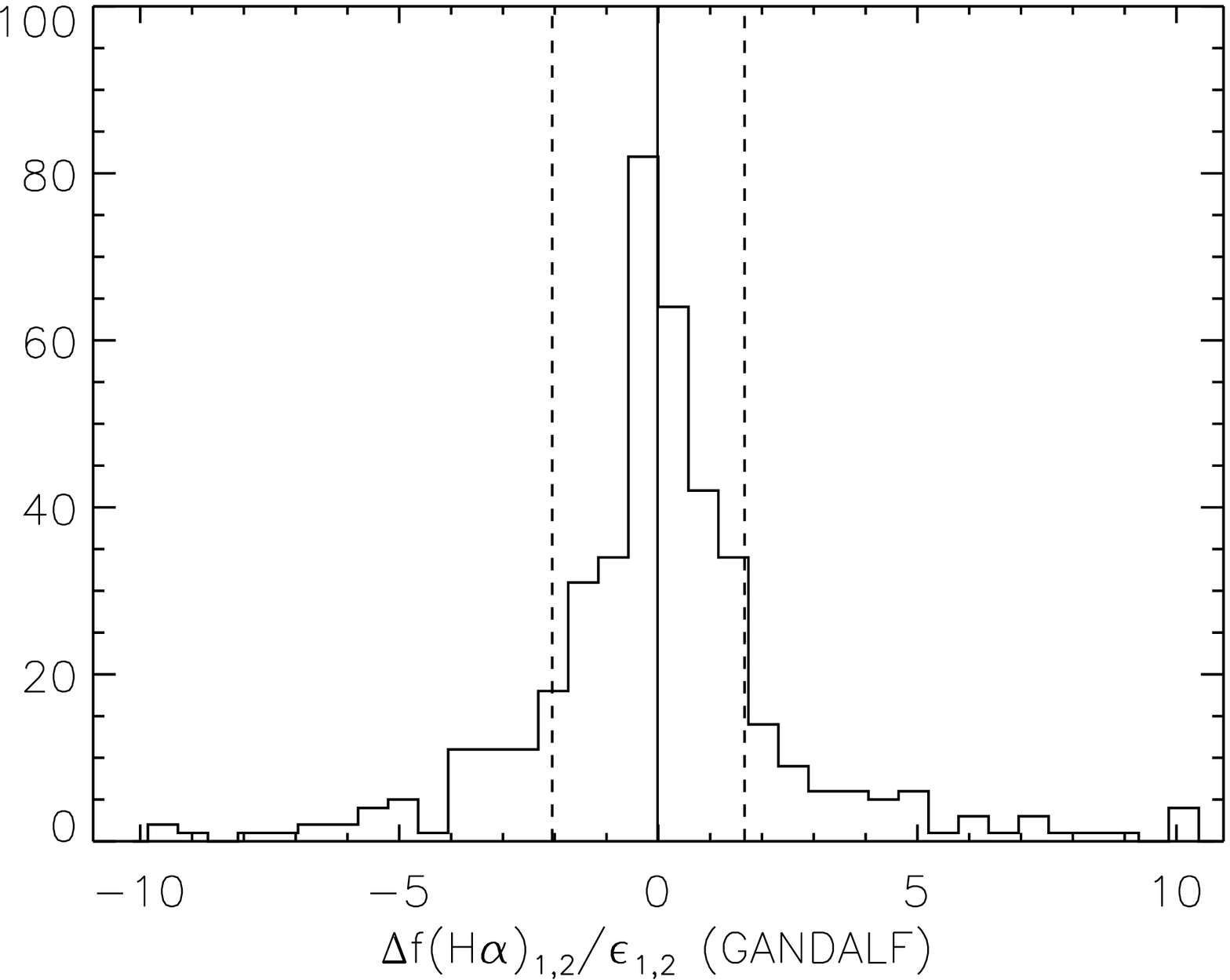}}\vspace{2mm}
\centerline{\includegraphics[width=60mm]{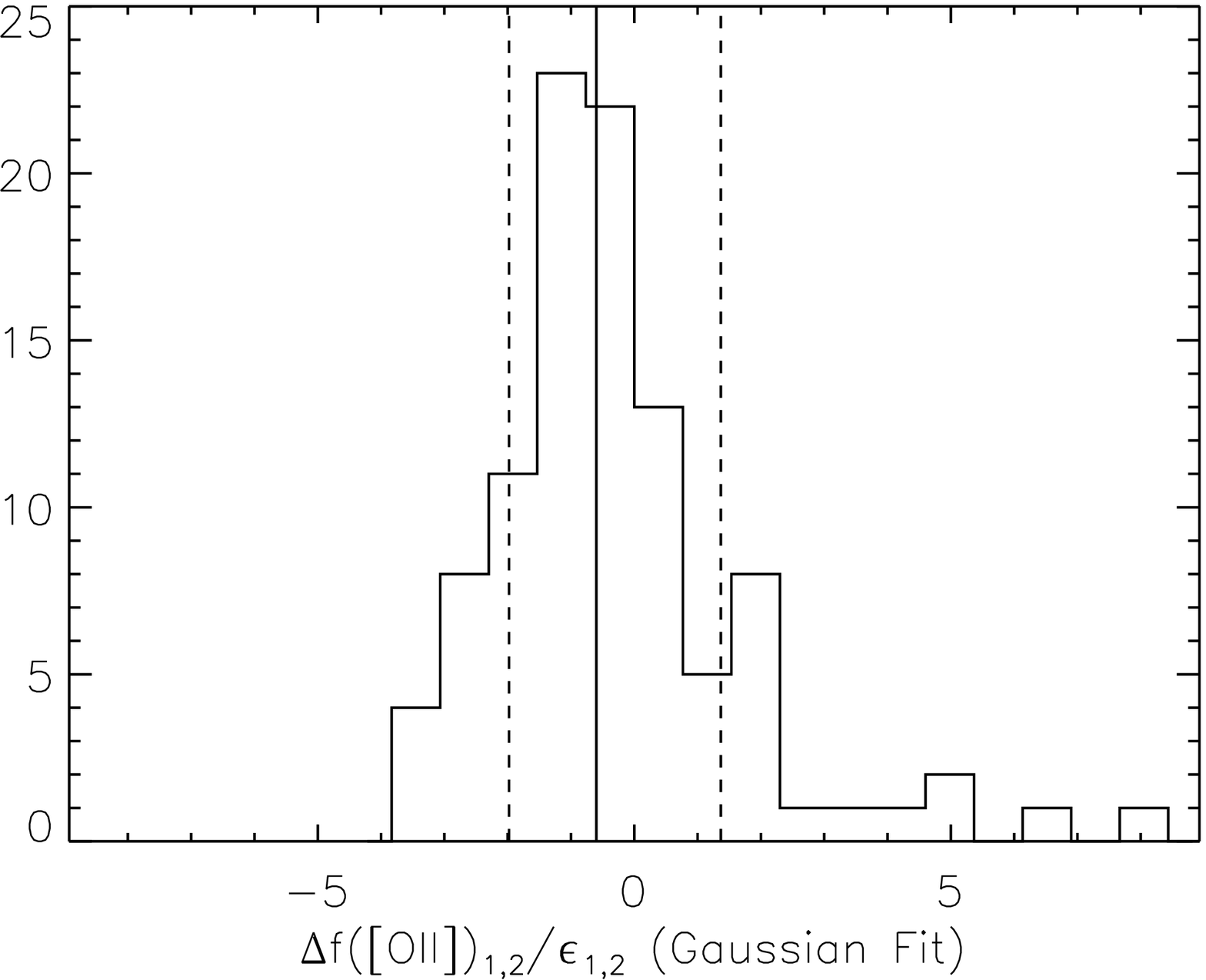}\hspace{1cm}
\includegraphics[width=60mm]{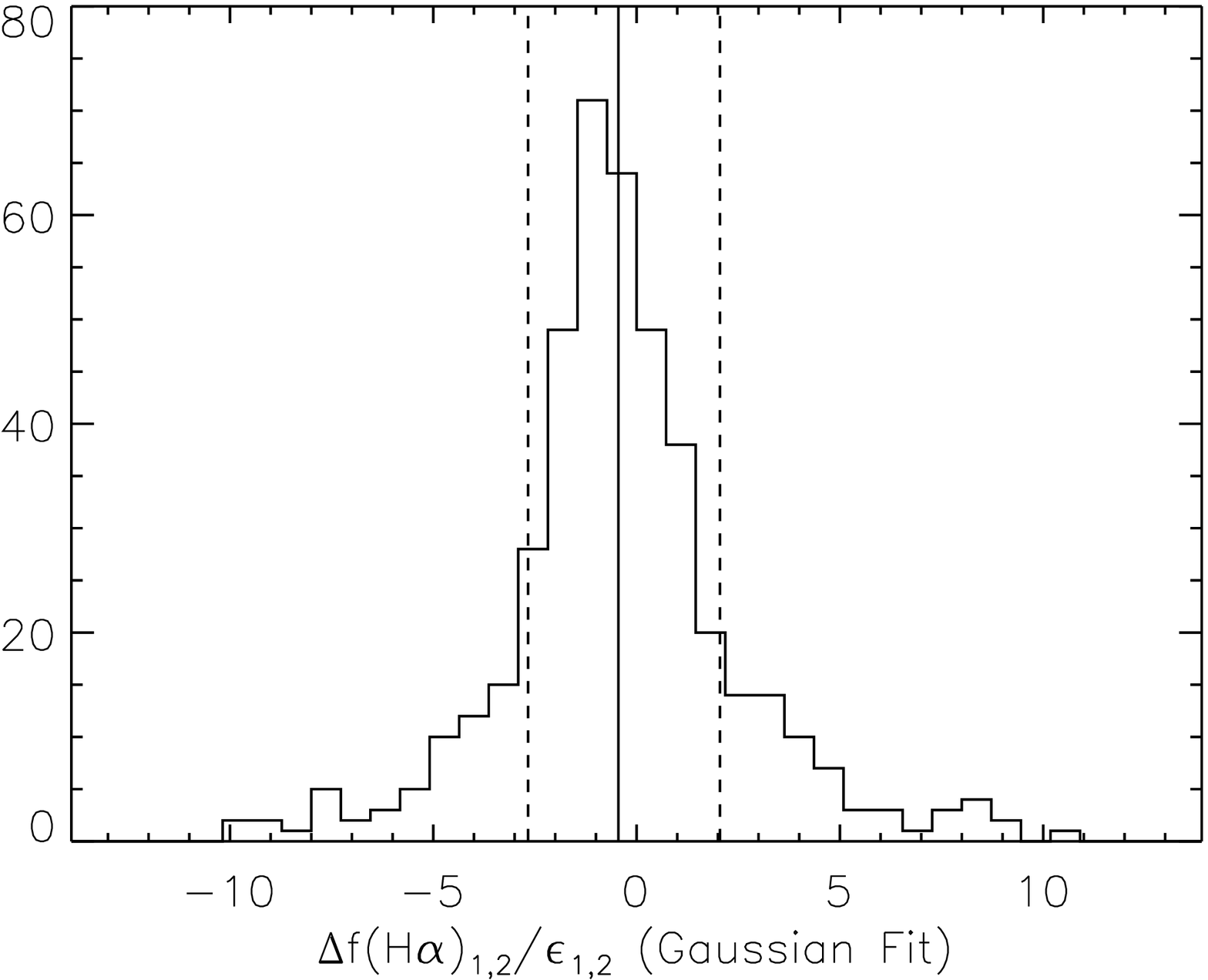}}
\caption{Same as Fig.~\ref{fig:F_dup_comp} but the difference is normalised by $\epsilon_{1,2}=\sqrt{(\sigma_1^2+\sigma_2^2)}$. If no biases in the measurements exist due to differences in the spectrum $S/N$ ratio, then the distributions should be centred at zero and be symmetric. If the uncertainty estimates are accurate measures of the true uncertainties, then the 68th percentile values should be of order unity.
\label{fig:F_rel_dup_comp}}
\end{figure*}

\begin{figure*}
\centerline{\includegraphics[width=60mm]{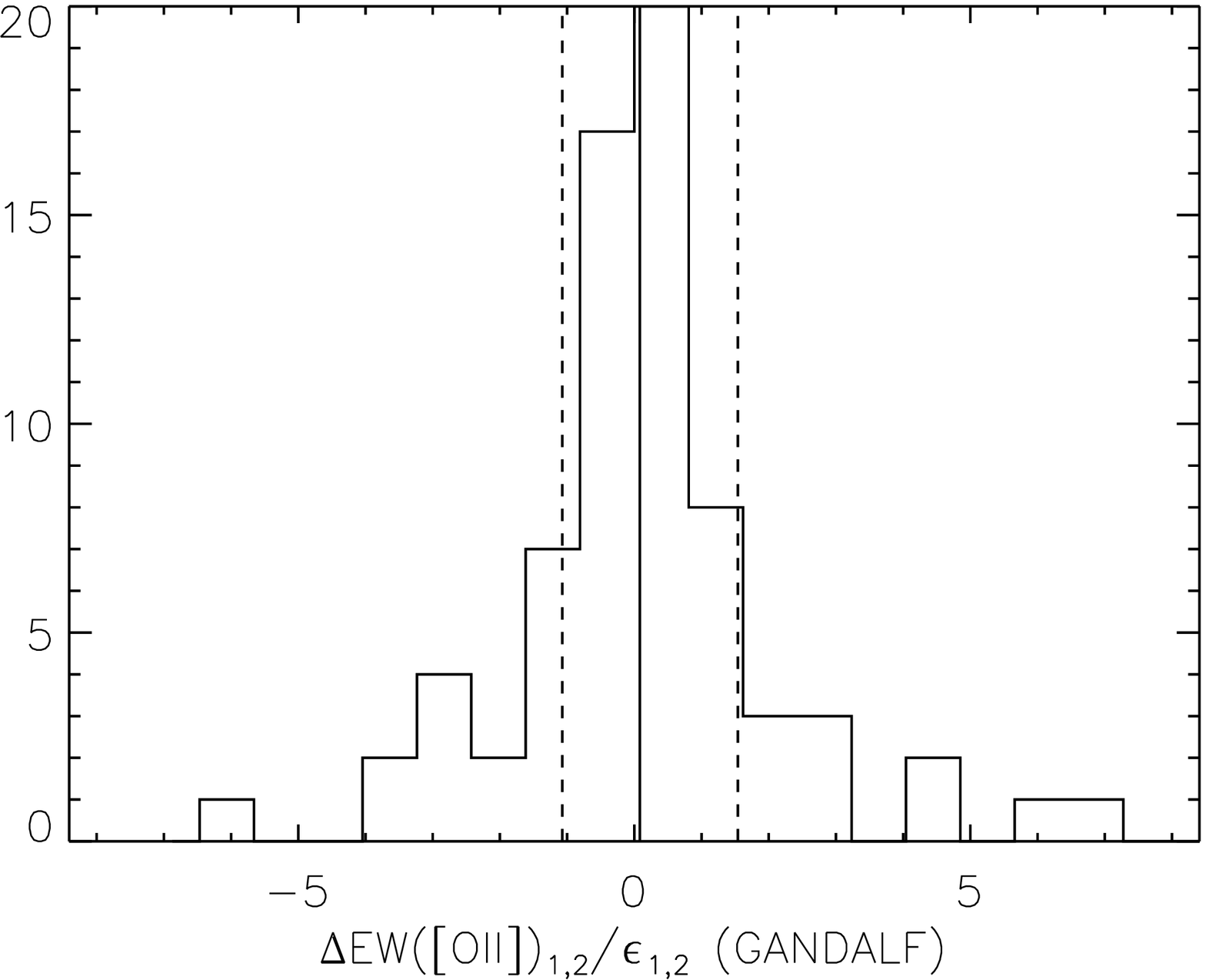}\hspace{1cm}
\includegraphics[width=60mm]{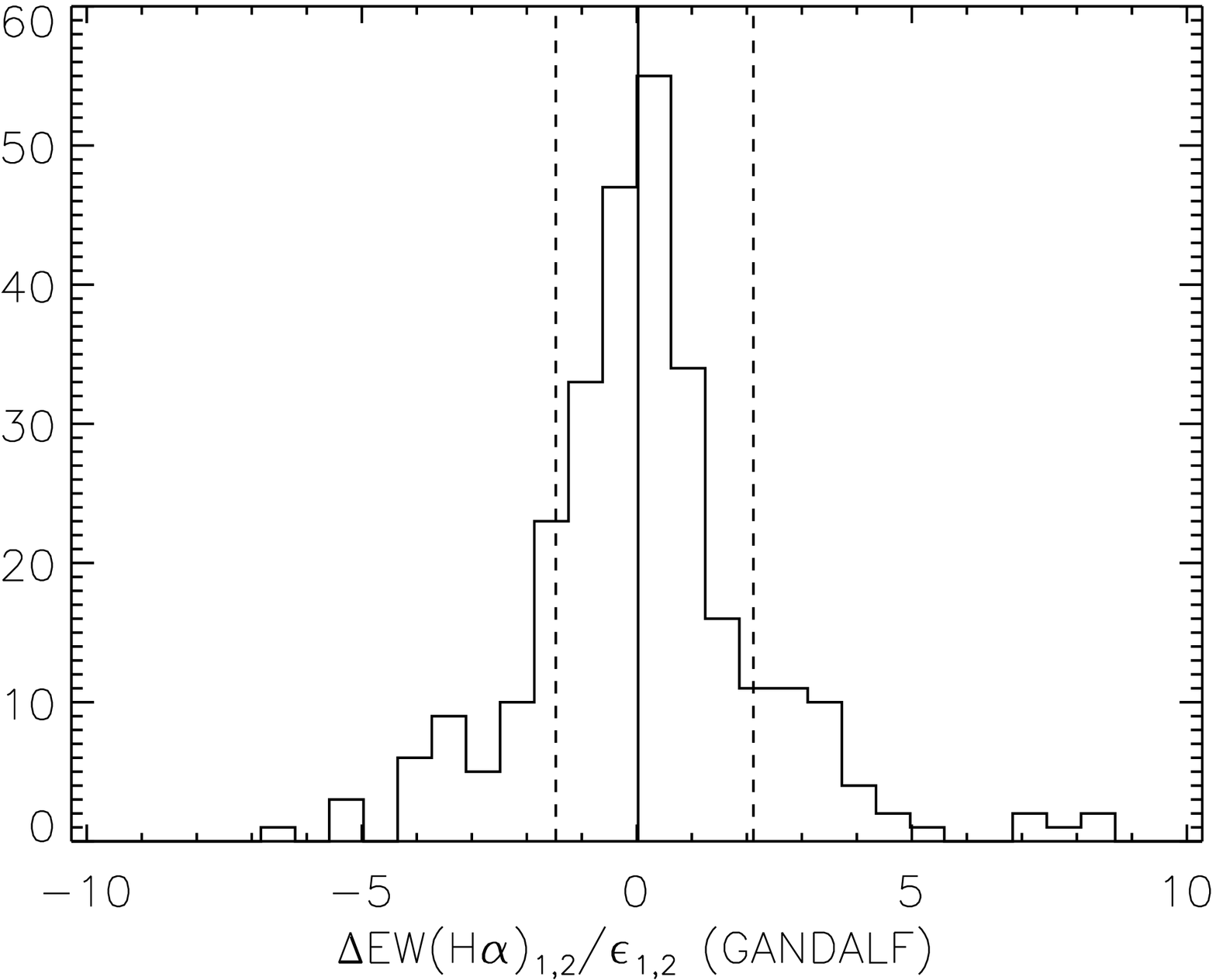}}\vspace{2mm}
\centerline{\includegraphics[width=60mm]{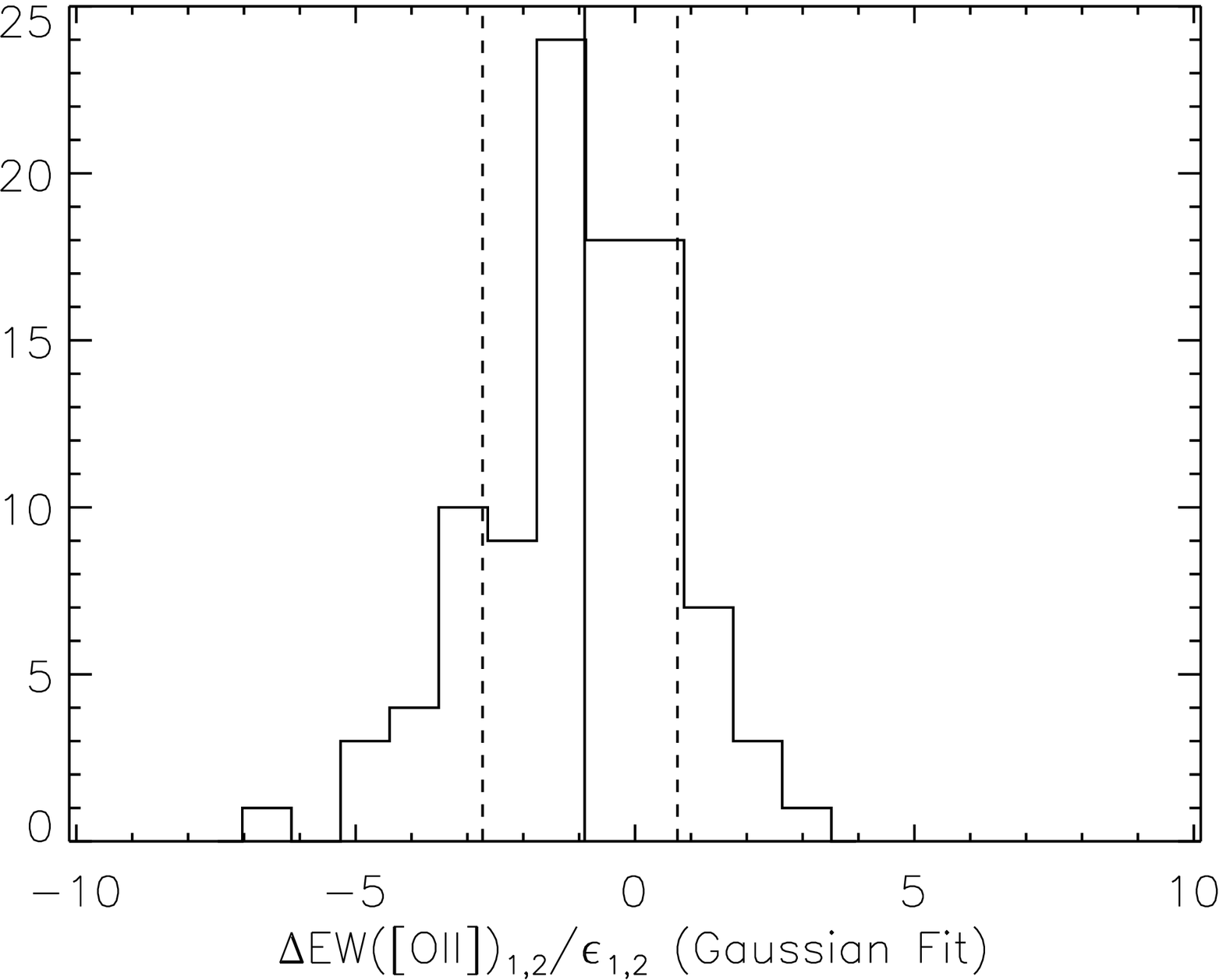}\hspace{1cm}
\includegraphics[width=60mm]{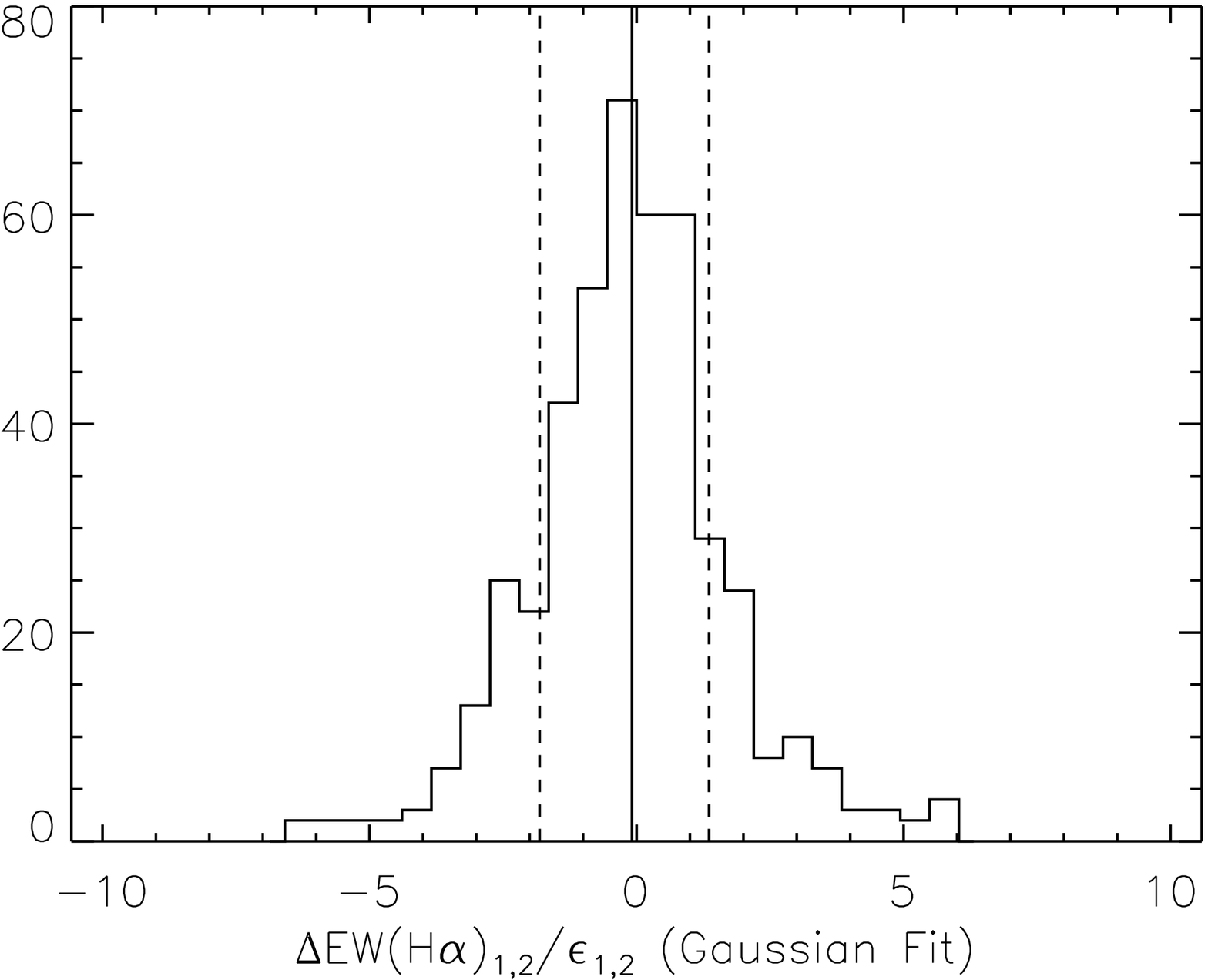}}
\caption{Same as Fig.~\ref{fig:F_rel_dup_comp} but for the line equivalent width measurements, rather than the fluxes.
\label{fig:ew_rel_dup_comp}}
\end{figure*}

\begin{figure*}
\includegraphics[width=0.9\textwidth]{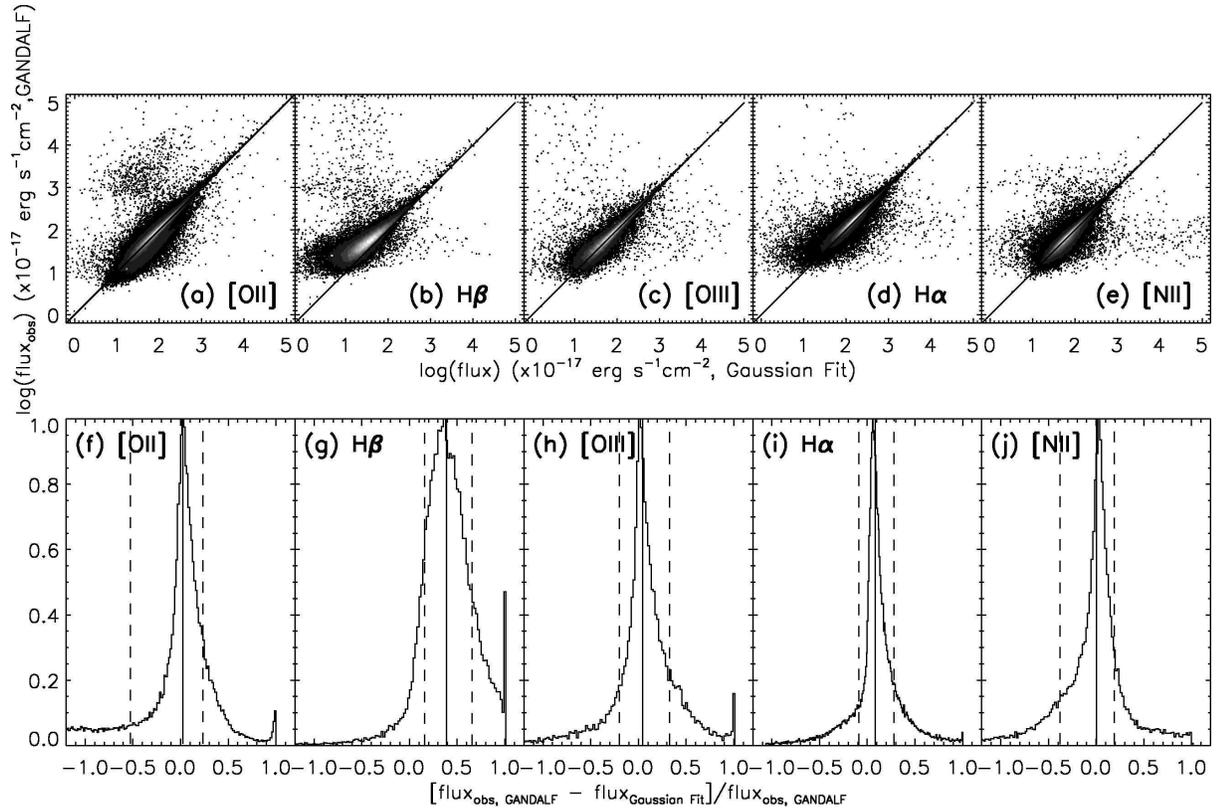}
\caption{Comparison of line flux measurements using GANDALF and Gaussian-fit methods. The top row shows the direct comparison of the two methods, with the deviation from the one-to-one relation for the Balmer lines being due to
stellar absorption. The bottom row shows a histogram of the differences in the measurements normalised by the GANDALF measurement. The tails illustrate cases where the two values are quite different, with
the spike visible at unity corresponding to cases where the Gaussian-fit flux is significantly smaller than the
GANDALF value. The tail to negative values corresponds to the GANDALF measurement being significantly
smaller than the Gaussian-fit value.
\label{fig:Flux_comp}}
\end{figure*}

\begin{figure*}
\includegraphics[width=0.9\textwidth]{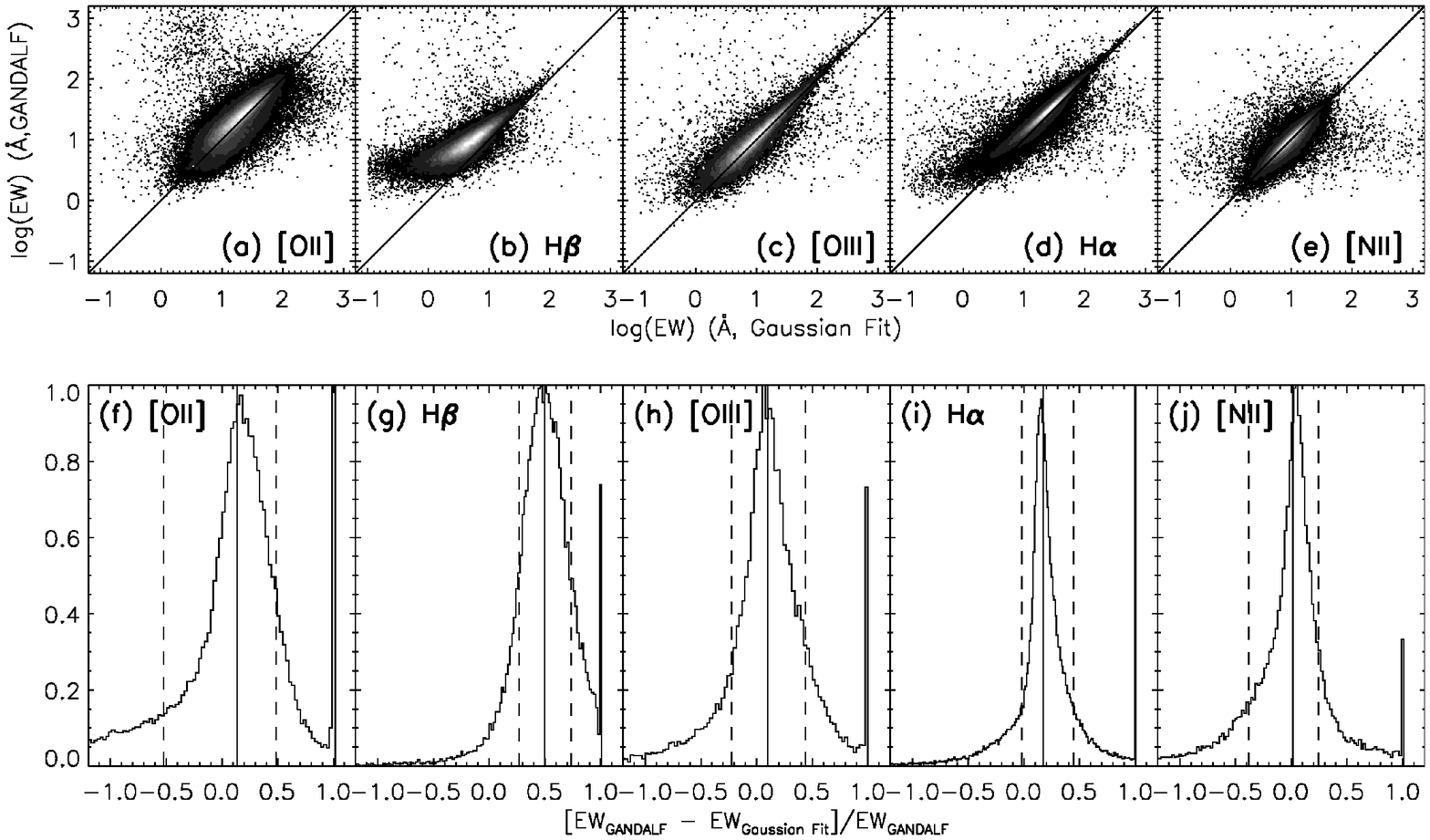}
\caption{Same as Fig.~\ref{fig:EW_comp}, but for line equivalent widths rather than fluxes.
\label{fig:EW_comp}}
\end{figure*}

\subsection{Internal consistency}
\label{intcon}
A simple test of the internal consistency of our line measurements is to measure the ratios
of emission lines from ionised species that should be fixed by quantum mechanics for fixed
density and temperature, with the
[OIII], [NII] and [SII] line pairs being obvious choices. Due to the stellar absorption of H$\alpha$
in the vicinity of the [NII]$\lambda6548$ line, and (to a lesser degree) its proximity to
the H$\alpha$ emission line itself, this particular ratio is less robust. Using the measurements
from the Gaussian fits, in Fig.~\ref{fig:lineratios}
we show the distribution of [OIII]$\lambda4956$/[OIII]$\lambda5007$ compared to the expected
ratio of $1/2.98$ \citep{SZ:00}. The same is shown for [SII]$\lambda6716$/[SII]$\lambda6731$, where
the expected ratio is $1/1.4$ \citep{Ost:89}. Line ratios are only included in this analysis for pairs where
the brighter of the two lines has a flux measurement above $3\,\sigma$.
In the right panels, the ratio is shown as a function of the flux of the brighter line.
At lower flux (more specifically lower $S/N$), the fainter line is less easily detected,
and a bias toward higher values of the ratio can be seen. This result demonstrates
the robustness of the line fitting and measurement within each spectrum.

\subsection{Duplicate measurements}
\label{dupobs}
A small number of GAMA spectra duplicate observations of particular targets, often due to the
re-observation of objects where a low redshift quality was initially obtained. This is typically a consequence
of the objects being at the fainter end of the GAMA target selection, and the results here are
consequently illustrative of the robustness of the measurements for the fainter population. There is no attempt
made to combine these duplicate GAMA spectra, or the measurements from them, due to the complex
systematics involved in the flux calibration steps detailed above. We can, however, take advantage
of these duplicate observations to understand the
precision to which we can measure our emission line equivalent widths and fluxes.
To do this, we select those duplicate measurements where both spectra have been allocated a
redshift quality $3 \le nQ \le 4$ \citep{Dri:11}. We illustrate the differences between these duplicate
measurements, using both the GANDALF and the Gaussian fits, for line fluxes (Fig.~\ref{fig:F_dup_comp})
and equivalent widths (Fig.~\ref{fig:ew_dup_comp}) of the H$\alpha$ and [OII] emission lines.
These two line species were chosen for the following reasons. First, they represent the best- and worst-case
scenarios with H$\alpha$ being (typically) a strong line in a high $S/N$ region of the spectrum, and [OII]
being both weak (in many cases) and situated at the low $S/N$ end of the spectrum. Thus, these two lines
give an idea of the spread expected in the precision of the measurements across the full wavelength
and $S/N$ range. Second, the ${\rm H}\alpha$ line is affected by underlying absorption due to the stellar
continuum which is corrected for during the GANDALF fitting process. However, this correction relies critically on a
robust model fit to the underlying stellar continuum and errors in this fit may add systematic uncertainties to the flux
and equivalent width measurements. Such systematics are quantified by comparing the differences in the
duplicate measurements to the quadrature sum of the uncertainties on the measurements
(Fig.~\ref{fig:F_rel_dup_comp} and Fig.~\ref{fig:ew_rel_dup_comp}). The 68th percentiles are typically
1.5 to 2 units from the median, suggesting that the formal statistical errors on the line measurements
are somewhat underestimated from both the GANDALF and the Gaussian fits. This is likely related
to the fact that the duplicate spectra are dominated by fainter galaxy targets. For the fainter or lower $S/N$
targets, the continuum level in the spectra tends to be noisier, and more affected by systematics such
as poor scattered light subtraction or sky subtraction. This leads in turn to greater variation between
the repeat measurements, as the continuum level estimated for either the Gaussian fitting or by GANDALF
can be more easily over- or under-estimated. The errors on the line fitting may not accurately reflect the
uncertainty in the continuum estimation.
For brighter (higher $S/N$) targets, this is less likely to be a limitation for the Gaussian fits, although
systematic underestimates in the uncertainties may still be possible in the GANDALF fitting if the continuum
is not well-described by the underlying SEDs.
Overall, the repeatability of the duplicate measurements is very high, with $1\,\sigma$ differences
of less than $0.5\,$\AA\ for the equivalent width of H$\alpha$ and less than $1\,$\AA\ for [OII]. The
repeatability of the flux measurements is similarly reliable, with $1\,\sigma$ variations of typically less than
$0.5\times10^{-17}\,$erg\,s$^{-1}$\,cm$^{-2}$.

\subsection{Self-consistency}
\label{selfcon}
Having established the precision of the emission line measurements in the GANDALF and Gaussian-fit catalogues
individually, using the duplicate measurements, we now test for systematic biases which may be inherent in the different
techniques used to determine the line equivalent widths and fluxes. To identify such potential biases, in
Fig.~\ref{fig:Flux_comp} and Fig.~\ref{fig:EW_comp} we investigate comparisons between the two measurement methods
for the equivalent widths and fluxes, respectively, for the [OII], ${\rm H}\beta$, [OIII]\,$\lambda 5007$, ${\rm H}\alpha$
and [NII]\,$\lambda 6584$ line species. In these Figures, we only include comparisons where the emission line of
interest is detected with an amplitude-to-noise ratio $A/rN > 3$ as determined for the GANDALF fits.
Here $A/rN$ is the ratio of the line amplitude (from the Gaussian fit) to the standard deviation of the residual spectrum
\citep{Sar:06}.
We also limit ourselves to cases where both line measurements have an equivalent width greater than
zero (noting that we take the convention that emission lines have positive equivalent widths).
The forbidden lines, [OII], [OIII]\,$\lambda 5007$ and [NII]\,$\lambda 6584$ are
chosen because they are not significantly affected by stellar absorption and probe a significant portion of the
wavelength range covered by the AAOmega spectra. Since the GANDALF measurements correct for the effects of
stellar absorption on the emission line measurements, the comparison between the Balmer line species,
${\rm H}\beta$ and ${\rm H}\alpha$, allows us to quantify the systematic effects of the underlying stellar absorption
on the Gaussian-fit measurements.

Considering the forbidden lines only, the distributions are consistent with the one-to-one relation. The median
and modes of the relative difference in the fluxes for [OII] and [OIII]\,$\lambda 5007$ (lower panels,
Fig.~\ref{fig:Flux_comp}) indicate offsets less than $5\%$ (with a slight systematic toward the GANDALF
measurements being larger). For the equivalent widths of [OII] and [OIII]\,$\lambda 5007$, however, the median
and modes of the relative differences (lower panels, Fig.~\ref{fig:EW_comp}) indicate that the GANDALF values are systematically higher by around $10-15\%$. The likely cause of this small offset in the equivalent width measurements
is in the different definitions used for the continuum flux estimates. The [NII] line, though, does not appear to
be affected by an offset of the same magnitude (equivalent width differences of at most 5\%), which further
suggests that the bulk of the difference arises in the continuum estimate for the noisier blue arm of the spectra
([OII] is always in the blue for GAMA spectra, while [OIII] is in the blue for spectra with $z<0.13$). The conclusion
here is that the independent measurement of emission line fluxes are consistent in the median to better than 5\%, with a
dispersion consistent with the error measurements on the lines. The line EW estimates are also consistent
to better than 5\% in the red, and to $10-15\%$ in the blue, again with dispersions consistent with the measured errors.

\subsection{Stellar absorption}
\label{stelabs}
\begin{figure}
\centerline{\includegraphics[width=80mm]{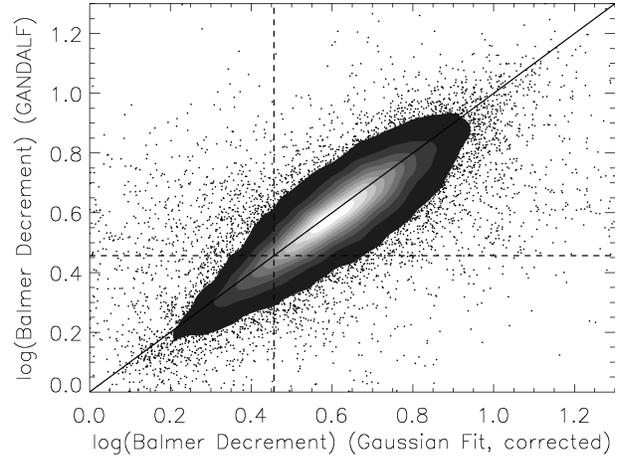}}
\caption{Balmer decrement ($F_{H\alpha}/F_{H\beta}$) as measured by GANDALF, compared against
that from the Gaussian line fits after correcting for a constant stellar absorption equivalent width of 2.5\,\AA. The
solid line indicates equality, and the dashed lines correspond to the value of 2.86 expected
for Case B recombination.
\label{fig:BD_comp}}
\end{figure}

As expected for the Balmer lines, the GANDALF flux and equivalent width measurements are systematically
higher than the Gaussian fit measurements due to stellar absorption, which is accounted for by
GANDALF but not in the Gaussian fits. The difference is particularly conspicuous for the
weaker H$\beta$ emission line where the stellar absorption can be large compared to the typical
H$\beta$ emission strength. The relative difference in the equivalent width measurements for H$\beta$
is offset by $\sim 50\%$, while the offset is $\sim 15\%$ for ${\rm H}\alpha$. The offset is dominated by
the correction to the underlying stellar absorption in the GANDALF measurements, although the
systematic offsets affecting the equivalent width measurements for [OII] and [OIII] may also be present at some level.
We find that applying an average stellar absorption correction to
the equivalent widths of $2.5\,$\AA\ is appropriate in order to make the Gaussian line fits consistent
with those from the GANDALF measurements. This can be seen explicitly in a comparison of the
Balmer decrements from the GANDALF measurements compared against the stellar-absorption-corrected
Gaussian fits in Fig.~\ref{fig:BD_comp}. This figure also indicates the typical Case B recombination
value of 2.86 for the Balmer decrement \citep{Ost:89}, assuming $T_e=10000\,$K and $n_e=100\,$cm$^{-2}$.
Note, though, that the intrinsic Balmer decrement can be as high as $\sim3$ for temperatures
$T_e\approx 5000-6000\,$K \citep[e.g.,][]{LSE:09}.
The inferred stellar absorption correction of $2.5\,$\AA\ is somewhat smaller (by $\approx 1\,$\AA) than the
typical stellar absorption equivalent widths for disk galaxies \citep{Ken:83}, but consistent with the
result found by \citet{Hop:03}. This is also consistent with stellar absorption equivalent widths, from
$1-2.5\,$\AA, found in a sample of star forming galaxies by \citet{LSE:10}, who show that the stellar
absorption tends to increase with increasing stellar mass, or metallicity, of a galaxy.
When the stellar absorption is marginally resolved, as it was
for the SDSS spectra analysed by \citet{Hop:03}, and as it is for the current AAOmega spectra,
a Gaussian fit to the line flux requires a smaller correction.

\begin{figure}
\centerline{\includegraphics[width=80mm]{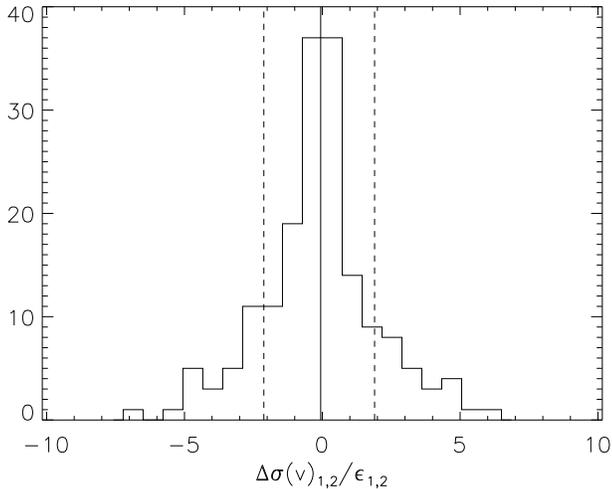}}
\caption{Differences between repeat measurements of the velocity dispersions, normalised by the quadrature
sum of the errors. This analysis is restricted to duplicate spectra where both of the pair have $S/N>3$ in the blue.
The 68th percentile of this distribution is about $\pm2$, suggesting that the errors in the velocity dispersions
in these relatively low $S/N$ spectra are underestimated by a factor of two.
\label{fig:veldisp}}
\end{figure}

\begin{figure}
\centerline{\includegraphics[width=80mm]{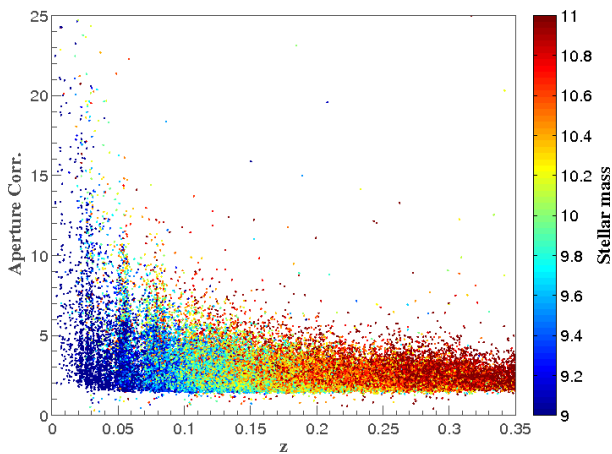}}
\caption{Effective aperture correction as a function of redshift. The correction is a multiplicative
flux scaling. The colour coding indicates the stellar masses of the galaxies.
The majority of GAMA systems have aperture corrections
of $2-4$, with a small number of galaxies at the lowest redshifts
having larger corrections.
\label{fig:apcor_z}}
\end{figure}

\begin{figure}
\centerline{\includegraphics[width=80mm]{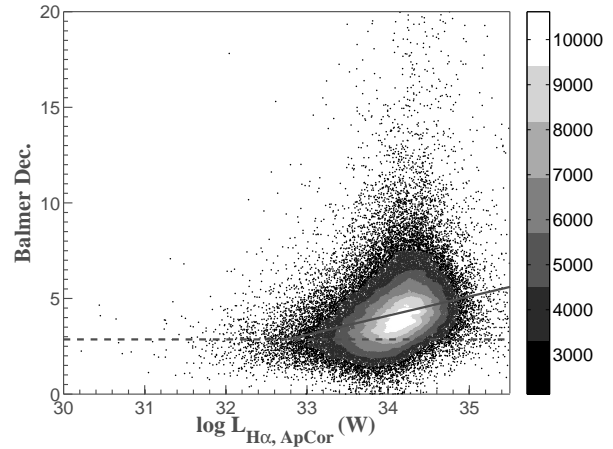}}
\caption{Balmer decrement as a function of aperture corrected (but not obscuration corrected)
H$\alpha$ luminosity. Systems classified as AGN are not shown (see \S\,\ref{spec_diag} below).
The dashed line shows the Case B recombination value of 2.86, and the solid line shows a fit to
the observations with Balmer decrements higher than this value.
\label{fig:bd}}
\end{figure}

\subsection{Velocity dispersions}
\label{vdisp}
Velocity dispersions, as measured by pPXF, are reliable for relatively high signal-to-noise
spectra. Earlier work suggests a conservatively high threshold of $S/N>12$ \citep{Pro:08}, for
extracting reliable velocity dispersions, but more recent work suggests that spectra with
$S/N>5$ \citep[Thomas et al., in prep.]{Shu:11} may still yield reliable measurements.
Fig.~\ref{fig:veldisp} shows the repeatability of the velocity
dispersion measurements from duplicate observations. The 68th percentile of this distribution is
around two, implying that the measured velocity dispersion errors are underestimated by about
a factor of two. This is likely a consequence of the relatively low $S/N$ of the duplicate spectra available
to make this measurement, though, especially in the blue (Fig.~\ref{fig:snmag}), since the absorption
features used in constraining the velocity dispersion lie primarily in the blue half of the spectrum.
In particular, errors in the estimate of the continuum level are again likely to
be contributing substantially to the underestimate in the errors on the velocity dispersions, for the
low $S/N$ spectra.
We have used duplicate spectra with $S/N>3$ in
order to sample enough duplicates for this analysis, since limiting the analysis only to $S/N>12$ rejects
the majority of the duplicate spectra. Consequently, since the velocity dispersion measurements
rely largely on absorption lines at the blue end of the spectra, these low $S/N$ duplicate spectra are likely to
be less well characterised than those of higher $S/N$.

\begin{figure*}
\centerline{\includegraphics[width=\textwidth]{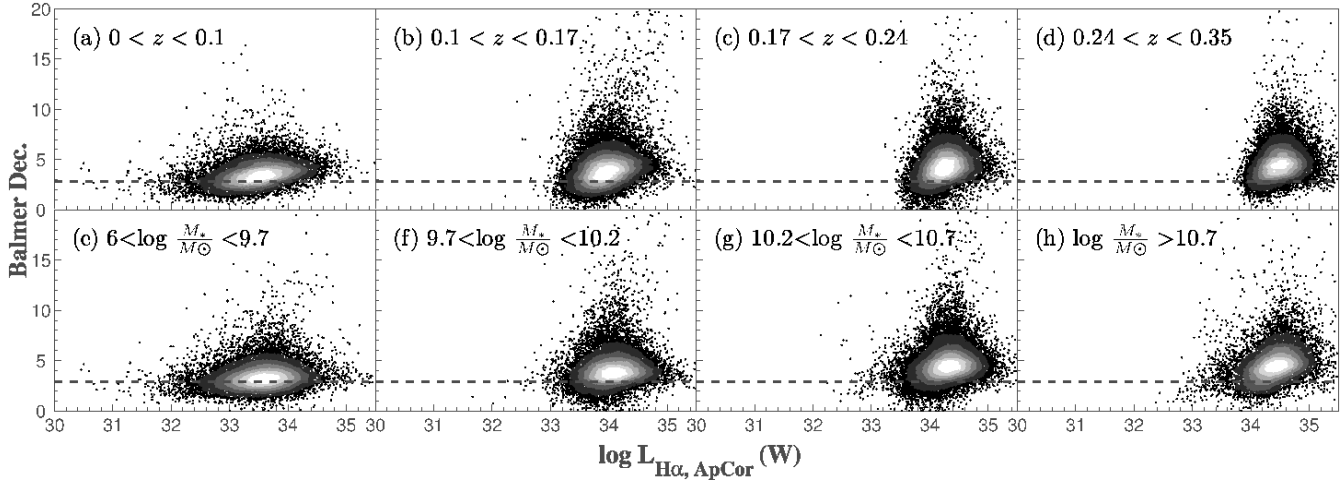}}
\caption{Balmer decrement as a function of aperture corrected (but not obscuration corrected)
H$\alpha$ luminosity. The top row shows the diagram separated into bins of redshift,
the bottom row into bins of galaxy stellar mass, both increasing left to right. The dashed line
is again the Case B value of 2.86.
\label{fig:bd_vs_mz}}
\end{figure*}

\subsection{Aperture effects}
\label{apeff}
Due to the $2''$ fibre diameter, emission from a galaxy that is larger than this on the sky
will not be measured within the fibre \citep[e.g.,][]{Hop:03,Bri:04}. A detailed analysis
of the systematic effects in fibre-spectrograph measurements associated with fibre-positioning errors,
efficiencies and aperture corrections is given by \citet{New:02}. Below we outline the extent of
the aperture effects in the GAMA spectra.

Corrections to account for these aperture losses
are incorporated into the flux calibration process above ({\S\,\ref{corrections}), and
are effectively multiplicative flux scalings. These can only ever be approximations,
of course, making the assumption that the spectroscopic properties (such as line fluxes, or
derived properties such as star formation rate) that lie outside the fibre aperture can
be scaled by the broadband light profile available from the photometry.
\citet{Kew:05} highlight, for example, how aperture
effects can bias the estimate of star formation rate, nebular metallicity and obscuration in galaxies.
\citet{Ger:12} presents an analysis highlighting the potential biases involved, for a sample of galaxies at $z<0.1$,
in particular comparing estimates of star formation rate based on integral field data to those
from aperture-corrected single-fibre measurements.
They emphasise that a simple aperture correction of the kind applied here can underestimate
the true effect by factors of $\sim 2.5$ on average, although with a large scatter.
Circumventing
the limitations of such a fixed fibre covering fraction can only truly be achieved with multiobject
integral field spectroscopy, promised by the next generation of instrumentation
such as the SAMI \citep{Croom:12} instrument on the AAT. 
There is no simple
systematic solution, because of the underlying variety of differently distributed star formation
locations within galaxies. In the absence of such detailed measurements, a simple aperture
correction still remains the best proxy for a total line luminosity or star formation rate for
a galaxy, and we briefly discuss here the scale of these corrections for the GAMA spectra.

To illustrate the extent of the aperture corrections, Fig.~\ref{fig:apcor_z} shows the aperture
correction as a function of redshift. This can be compared against aperture corrections using a
similar approach, seen in Fig.~25(a) of \citet{Hop:03}, which shows the same quantity (given
logarithmically) as a function of redshift for SDSS spectra. For the SDSS galaxies,
the aperture corrections vary from factors of $\sim 2$ to 10 or more at the lowest redshifts. Comparing
the SDSS galaxies with the GAMA targets, a larger fraction of the SDSS systems
below $z\sim 0.1$ have aperture corrections larger than a factor of 5, due to the larger size
on the sky of the low-redshift galaxies, despite the larger SDSS fibre aperture.

The GAMA spectra have been taken using $2''$ diameter fibres, compared to the $3''$ diameter fibres
used by SDSS. This is not as problematic as might be initially assumed, however, as the GAMA
targets (primarily $17.77<r<19.8$) are fainter than those from SDSS ($r<17.77$).
Consequently, at a fixed redshift the typical GAMA target is significantly smaller (in kpc) than
the typical SDSS galaxy. Similarly, at fixed mass, the typical GAMA target is at higher redshift, and so
smaller (in arcsec). It turns out that the distribution of aperture sizes in units of the effective radius are
very similar between GAMA and SDSS. Thus, as a consequence of the fainter magnitude limit, and
greater redshift depth, despite the fact that 2dF/AAOmega has smaller fibres than
SDSS, we are no more susceptible to aperture effects.

\section{Emission line ratios}
Ratios of bright emission line measurements for galaxies are commonly used to
constrain the obscuration (through the Balmer decrement, H$\alpha$/H$\beta$), or
to discriminate between a supermassive black hole (active galactic nucleus, AGN)
or star formation (SF) as the photoionisation source. A typical spectroscopic
diagnostic use for the latter discrimination compares the [O{\sc iii}]/H$\beta$ to
[N{\sc ii}]/H$\alpha$ ratios, following \citet{BPT:81}. In this section we demonstrate the range
of such properties present in the GAMA sample. Here we show results using the
Gaussian fits to the emission lines, although we see identical trends if we use
the GANDALF measurements.

\begin{figure}
\centerline{\includegraphics[width=80mm]{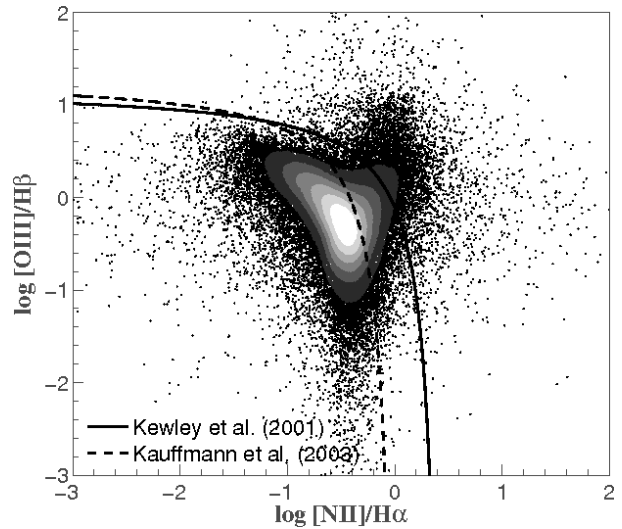}}
\caption{Spectral diagnostic diagram demonstrating the range of diagnostic measurements in the
GAMA sample. The discrimination lines shown
are from \citet{Kew:01} (solid) and \citet{Kau:03b} (dashed).
\label{fig:bpt}}
\end{figure}

\subsection{Balmer decrement}
The Balmer decrement is calculated as the ratio of stellar-absorption corrected line flux, H$\alpha$/H$\beta$.
The distribution of Balmer decrement, as a function of stellar absorption corrected and aperture
corrected H$\alpha$ luminosity, is shown in Fig.~\ref{fig:bd} \citep[see also][]{Gun:13}. Galaxies are only
shown in this figure if they have $S_{\rm H\alpha}>25\times10^{-17}$\,erg\,s$^{-1}$\,cm$^{-2}$, and local
volume flow corrected redshifts in the range $0.001<z<0.35$. The flux limit corresponds to the
limit at which the emission lines are robustly measured \citep{Gun:13}. The lower
redshift limit excludes galaxies with erroneously low redshift measurements or
stellar contaminants \citep{Bal:12}. The upper redshift limit is where H$\alpha$ falls outside the
observable spectral range. The trend demonstrates that
high line luminosity systems show a much broader distribution of obscuration properties than
lower luminosity galaxies, consistent with earlier results
\citep{Hop:01,Hop:03,Per:03,Afo:03,LS:10,Gun:11}.

With the numbers of galaxies available to us in the
GAMA sample, we can explore the distribution of obscuration properties in more depth by
looking at trends with both mass and redshift. We find results that are consistent with,
and complementary to the early SDSS work by \citet{Kau:03a} as well as many other recent works.
Fig.~\ref{fig:bd_vs_mz} demonstrates that
the low stellar mass galaxies tend typically to have lower H$\alpha$ luminosities,
and low overall levels of obscuration. Progressively higher mass systems, which can sustain higher
levels of star formation, tend to display both higher levels of H$\alpha$ luminosity and
more extreme Balmer decrements. Note that Fig.~\ref{fig:bd_vs_mz} shows the H$\alpha$
luminosity before correcting for obscuration, so the difference in intrinsic luminosities between
low and high mass systems will be enhanced. Interestingly, at the highest masses, there is a detectable
division into high H$\alpha$ luminosity, high obscuration systems, and lower H$\alpha$
luminosity, low obscuration systems. These high-mass, lower luminosity systems have low specific
star formation rates, and given their low obscurations are likely to correspond to
objects undergoing the transition from the blue cloud to the red sequence
\citep[e.g.,][]{Bal:04}. This population is explored in more detail in Taylor et al.\ (in prep).

When these trends are explored as a function of redshift the same broad picture emerges,
a consequence of the apparent magnitude limit of GAMA resulting in high-mass systems being preferentially
identified at higher redshift. There is a detectable difference at higher redshift, with the absence of
high-mass systems displaying low luminosities and obscurations. This is likely to be a
consequence of Malmquist bias, with such low-luminosity systems not being able to be
detected at higher redshift.

The interesting point to highlight here is that there is a bivariate selection effect at work
when exploring the properties of emission lines in a spectroscopic
survey of a broad-band magnitude-limited sample. The broad-band magnitude limit, to first order,
corresponds to a (redshift-dependent) stellar-mass limit. This is subsequently subjected to
an emission line flux limit through the spectroscopic observations, which, again to first order,
corresponds to a (redshift-dependent) luminosity limit. There will always be
galaxies that are bright enough in the continuum to be targeted spectroscopically, but whose
emission properties are too faint to detect. These can be accounted for with appropriate completeness
corrections. There will also exist galaxies that are brighter than the spectroscopic sensitivity
limits, but which never enter the spectroscopic sample as their hosts are fainter than the
broad-band magnitude limits used to select the targets. This component cannot be accounted
for with completeness corrections since their host population is not well-defined. 
The consequences of these bivariate selection effects in GAMA are discussed in detail in \citet{Gun:13}.

\begin{figure*}
\centerline{\includegraphics[width=\textwidth]{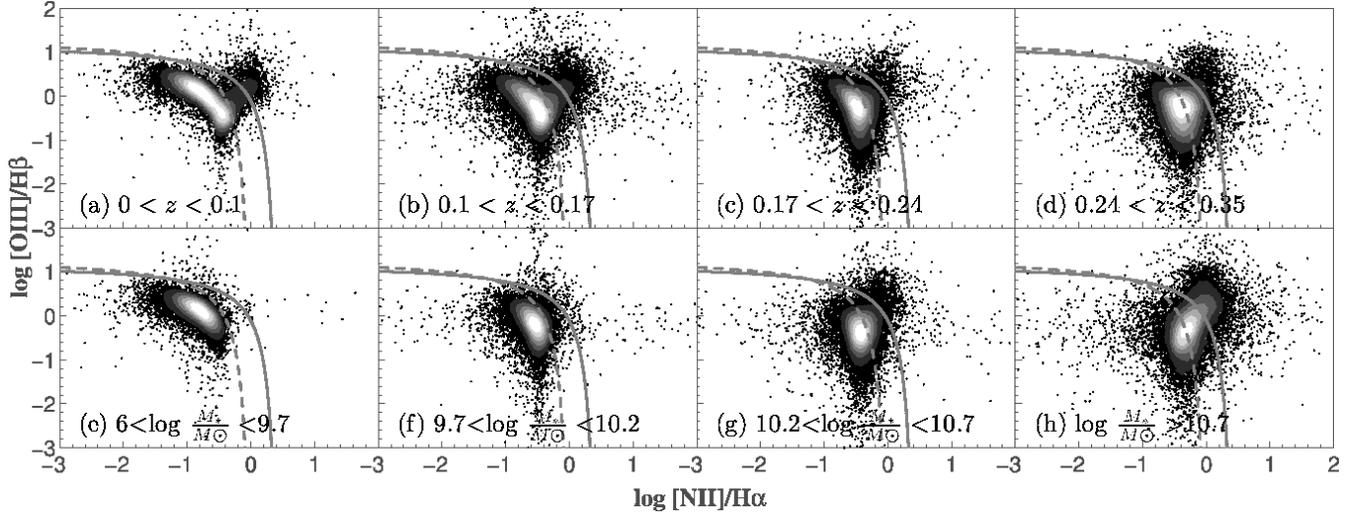}}
\caption{Spectral diagnostic diagram now illustrating the mass and redshift dependencies.
The top row shows the diagram separated into bins of redshift,
the bottom row into bins of galaxy stellar mass, both increasing left to right.
\label{fig:bpt_vs_mz}}
\end{figure*}

\subsection{Spectral diagnostics}
\label{spec_diag}

The ratios of forbidden emission lines to Balmer lines have been used for many
decades now as discriminators between different sources of photoionisation
\citep{BPT:81,VO:87,Kew:01}. In Fig.~\ref{fig:bpt} we show the most commonly
used diagnostic diagram, [OIII]/H$\beta$ as a function of [NII]/H$\alpha$, for
the full GAMA sample. Systems shown here are again limited to
those with $S_{\rm H\alpha}>25\times10^{-17}$\,erg\,s$^{-1}$\,cm$^{-2}$,
and local volume flow corrected redshifts in the range $0.001<z<0.35$.
This Figure discriminates star forming galaxies (below
the dashed line) from those where the ionisation arises from an AGN (above the solid line),
with galaxies between these discriminators commonly treated as composite systems.
Of the emission line systems in GAMA, the majority are classified as star forming
in this fashion. Only about 12\% of the galaxies with measured [NII], [OIII], H$\alpha$
and H$\beta$ fluxes, and quality ${\rm nQ}\ge3$, are classified as AGN.

Again capitalising on the sample size available with GAMA, we explore the mass
and redshift dependencies of the star formation and AGN distributions, shown
in Fig.~\ref{fig:bpt_vs_mz}. Perhaps not too surprisingly, the trends visible here
highlight that AGN systems are more prevalent in more massive galaxies, and
given the magnitude-limited nature of the sample, these are more visible at
higher redshift. At the lowest masses, the galaxy population is almost completely dominated
by star forming systems. As galaxy stellar mass increases, there is a progressive increase
in the proportion of AGN systems. These results are consistent with those demonstrated
from the SDSS \citep{Kau:03b,Hao:05}.

It is also worth noting the morphology of the region populated by the star forming
galaxies in Fig.~\ref{fig:bpt_vs_mz}. The star forming population moves from
a region of low [NII]/H$\alpha$ and high [OIII]/H$\beta$ for low mass systems, corresponding
to low metallicities, progressively
to having high values of [NII]/H$\alpha$ and low values of [OIII]/H$\beta$ for galaxies of
high stellar mass, corresponding to high metallicities. This transition reflects the
well-established mass-metallicity relationship for galaxies \citep[e.g.,][]{Tre:04,KE:08,LL:10}.
Details of the metallicity properties of galaxies in the GAMA sample are presented
by \citet{Fos:12} and \citet{LL:13}.

The redshift dependencies of this spectral diagnostic are also illuminating. As with
the Balmer decrements, in a broad sense the redshift trends reflect the mass dependencies
due to the magnitude-limited nature of the survey, which leads to high-mass systems
preferentially being found at higher redshift. Interestingly, though, at the lowest redshift
there is a population of LINER-like AGN, with high [NII]/H$\alpha$ with low [OIII]/H$\beta$ \citep{Scha:07},
which are relatively high-mass systems. This ionisation signature may also be more characteristic
of shock-excitation than the more prevalent active nucleus driven ionisation in massive
galaxies \citep{SharpBH:10,Far:10,Ric:10,Ric:11}.

\section{Conclusion}
\label{conc}
We have detailed above the processes involved in compiling, processing,
calibrating and measuring the AAOmega spectra that underpin the GAMA multiwavelength survey.
Details of the spectroscopic flux calibration provided in \S\,\ref{corrections} show that we achieve a
precision of about $10-20$\%. The continuum $S/N$ in the spectra is higher in the red than in the
blue, being typically $\sim 10$ in the red and $\sim 5$ in the blue for the brightest targets,
and decreasing as expected for fainter targets. The spectroscopic measurement reliability
has been quantified in terms of internal consistency, repeatability and self-consistency between
independent approaches to the emission line measurements. These analyses demonstrate that
the various measurements give consistent results with robustly estimated uncertainties.
It is important to note that we provide both the relatively direct Gaussian fit measurements, as well as
those from the more sophisticated GANDALF fitting, as GAMA data products. We do this recognising
that there will be some spectra for which GANDALF is not able to make a reliable measurement
(such as a badly spliced spectrum where the SED fitting has failed), but for which reliable
Gaussian fits to the emission features can still be made. There are also likely to be a variety of
science cases where having a larger number of simple measurements is more valuable than
having a smaller number of more refined measurements, and vice-versa. To facilitate both
aspects, both sets of measurements are provided.

The GAMA survey has already produced a broad cross-section of insights into the properties
of galaxy evolution, and as the survey progresses it will continue to provide a unique and valuable
spectroscopic and multiwavelength resource for studies of galaxy formation and evolution for many
years to come. All the GAMA results that have been published to date are based only on data from GAMA I,
i.e., the 144\,deg$^2$ contained within the three Equatorial fields, G09, G12, G15, taken during observing
campaigns spanning 2008--2010. The GAMA survey has continued through 2010--2012 with additional
observations of the Equatorial fields, to expand the area and achieve a uniform survey depth
of $r_{\rm pet}<19.8$\,mag, as well as opening up two new Southern fields, G02 and G23. The goal is
to survey a total area of 280\,deg$^2$ to a uniform depth of $r_{\rm pet}<19.8$\,mag, resulting
in $\sim 300\,000$ galaxy spectra. To date, over 220\,000 spectra have been measured.

We illustrate the data quality and utility with a simple exploration of how obscuration in galaxies
varies with galaxy mass and redshift, using the Balmer decrement. We find, consistent with earlier
work, a luminosity-dependence in galaxy obscuration. This effect is seen both as a function of
mass and redshift, largely as a consequence of the magnitude-limited nature of the survey. We
do identify, though, a population of high mass, low H$\alpha$ luminosity systems, with relatively
low obscuration, that are likely to be systems transitioning from the blue cloud to the red sequence.
We also explore the mass and redshift dependence of the spectral diagnostic diagram, finding that
AGN systems are more prevalent in higher mass galaxies, which are more numerous at
higher redshifts in magnitude limited samples like GAMA. Higher mass systems are also
less likely to have lower [NII]/H$\alpha$ ratios, consistent with having higher nebular metallicities, and
reflecting the well-established mass-metallicity relationship for galaxies. Both of these
results and more are being explored in more detail in a variety of works in progress.

The raw and processed GAMA spectra, and the derived data products, are being progressively
released to the public through staged Data Releases. The data and data products will be available from
the GAMA web site {\tt http://www.gama-survey.org/}. The GAMA team welcomes proposals from external
investigators interested in collaboratively using the dataset while it is still proprietary, by contacting the team
leaders at {\tt gama@gama-survey.org}.

\section*{Acknowledgements}

The authors warmly thank the referee, Michael Strauss, for extensive, detailed and considered
comments that have helped to improve this work.
Thanks to Jacob P.\ Crossett (Monash University) for contributions to the GAMA AAT observations in 2012.
AEB, MSO, MEC and MLL acknowledge support from the Australian Research Council (FS100100065, FS110200023).
JL acknowledges support from the Science and Technology Facilities Council (ST/I000976/1).
CF acknowledges co-funding under the Marie Curie Actions of the European Commission (FP7-COFUND).
PN acknowledges a Royal Society URF and ERC StG grant (DEGAS-259586).

Funding for the SDSS and SDSS-II has been provided by the Alfred P. Sloan Foundation, the Participating
Institutions, the National Science Foundation, the U.S. Department of Energy, the National Aeronautics and Space
Administration, the Japanese Monbukagakusho, the Max Planck Society, and the Higher Education Funding Council
for England. The SDSS Web Site is {\tt http://www.sdss.org/}.

The SDSS is managed by the Astrophysical Research Consortium for the Participating Institutions. The Participating
Institutions are the American Museum of Natural History, Astrophysical Institute Potsdam, University of Basel, University
of Cambridge, Case Western Reserve University, University of Chicago, Drexel University, Fermilab, the Institute for
Advanced Study, the Japan Participation Group, Johns Hopkins University, the Joint Institute for Nuclear Astrophysics, the
Kavli Institute for Particle Astrophysics and Cosmology, the Korean Scientist Group, the Chinese Academy of Sciences
(LAMOST), Los Alamos National Laboratory, the Max-Planck-Institute for Astronomy (MPIA), the Max-Planck-Institute for
Astrophysics (MPA), New Mexico State University, Ohio State University, University of Pittsburgh, University of
Portsmouth, Princeton University, the United States Naval Observatory, and the University of Washington.

GAMA is a joint European-Australasian project based around a spectroscopic campaign using the
Anglo-Australian Telescope. The GAMA input catalogue is based on data taken from the Sloan Digital
Sky Survey and the UKIRT Infrared Deep Sky Survey. Complementary imaging of the GAMA regions is
being obtained by a number of independent survey programs including GALEX MIS, VST KIDS, VISTA
VIKING, WISE, Herschel-ATLAS, GMRT and ASKAP providing UV to radio coverage. GAMA is funded
by the STFC (UK), the ARC (Australia), the AAO, and the participating institutions. The GAMA website
is {\tt http://www.gama-survey.org/}. 

\setlength{\bibhang}{2.0em}
\setlength\labelwidth{0.0em}

\label{lastpage}

\end{document}